\documentclass{amsart}

\usepackage{graphicx}
\usepackage{amsmath}
\usepackage{amssymb}
\usepackage{amsfonts}
\usepackage{amsaddr}
\usepackage{epstopdf}
\usepackage{color}
\usepackage{cite}

\newcommand{\sgn}{\text{sgn}}
\usepackage{float}
\usepackage[margin = 1in]{geometry}
\usepackage{hyperref}
\usepackage{caption}
\newcommand{\sech}{\text{sech}}
\usepackage[capitalise]{cleveref}
\crefformat{equation}{#2(#1)#3}

\title{Shock waves in dispersive hydrodynamics with non-convex dispersion}
\author{Patrick Sprenger, Mark A. Hoefer}
\email{patrick.sprenger@colorado.edu}

\address{Department of Applied Mathematics,
University of Colorado, Boulder, Colorado 80309-0526, USA}
\email{hoefer@colorado.edu}
\date{\today}
\begin{document}

\begin{abstract}
Dissipationless hydrodynamics regularized by
  dispersion describe a number of physical media including water
  waves, nonlinear optics, and Bose-Einstein condensates.  As in the
  classical theory of hyperbolic equations where a non-convex flux
  leads to non-classical solution structures, a non-convex linear
  dispersion relation provides an intriguing dispersive hydrodynamic
  analogue.  Here, the fifth order Korteweg-de Vries (KdV) equation,
  also known as the Kawahara equation, a classical model for shallow water waves, is shown to be a universal
  model of Eulerian hydrodynamics with higher order dispersive
  effects.  Utilizing asymptotic methods and numerical computations,
  this work classifies the long-time behavior of solutions for
  step-like initial data.  For convex dispersion, the result is a
  dispersive shock wave (DSW), qualitatively and quantitatively
  bearing close resemblance to the KdV DSW.  For non-convex
  dispersion, three distinct dynamic regimes are observed.  For small
  amplitude jumps, a perturbed KdV DSW with positive polarity and
  orientation is generated, accompanied by small amplitude radiation
  from an embedded solitary wave leading edge, termed a radiating DSW
  or RDSW.  For moderate jumps, a crossover regime is observed with
  waves propagating forward and backward from the sharp transition
  region.  For jumps exceeding a critical threshold, a new type of DSW
  is observed we term a translating DSW or TDSW.  The TDSW consists of
  a traveling wave that connects a partial, non-monotonic, negative
  solitary wave at the trailing edge to an interior nonlinear periodic
 wave.  Its speed, a generalized Rankine-Hugoniot jump condition, is
  determined by the far-field structure of the traveling wave. The
  TDSW is resolved at the leading edge by a harmonic wavepacket moving
  with the linear group velocity.  The non-classical TDSW exhibits
  features common to both dissipative and dispersive shock waves.
  \end{abstract}
\maketitle 
\section{Introduction}
\begin{figure}
\centering
\includegraphics[scale=0.25]{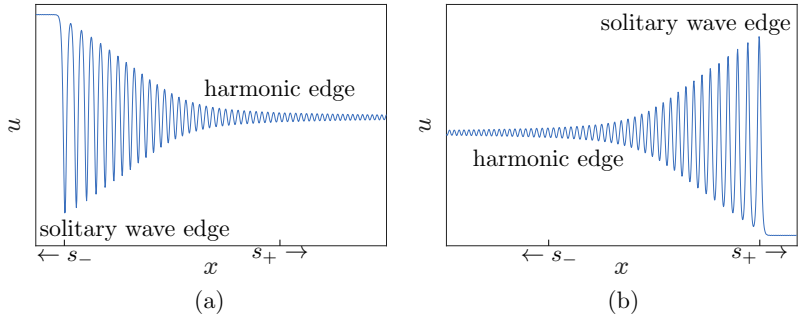}
\caption{Expanding KdV-like DSWs ($s_- < s_+$) for (a) $\sigma = -1$
  and (b) $\sigma = +1$, exhibiting dispersion curvature dependent
  orientation and polarity.}
  \label{fig:kdv_dsw}
\end{figure}

Dispersive hydrodynamics encompass hyperbolic systems of equations
regularized by dispersion rather than dissipation, modeling many
physical media \cite{el_dispersive_2016}.  One of the most prominent
features of these systems is a dispersive shock wave (DSW), in which
gradient catastrophe at the purely hyperbolic level is resolved into
an expanding, oscillatory wavetrain due to dispersive regularization.
The standard or classical DSW can be modeled by the Korteweg-de Vries
(KdV) equation
\begin{equation}
  \label{eq:1}
  u_t + uu_x + \sigma u_{xxx} = 0,
\end{equation}
where $\sigma = \pm 1$.  A KdV DSW can be described by a slowly
modulated, periodic traveling wave solution of \cref{eq:1} via Whitham
theory \cite{whitham2011linear}, exhibiting two distinguished edges
corresponding to a vanishing amplitude harmonic wavepacket and a
vanishing wavenumber solitary wave (see schematic DSWs in
\cref{fig:kdv_dsw}).  The canonical problem of interest is the
Gurevich-Pitaevskii (GP) problem, whereby the long time dynamics for
\cref{eq:1} with step initial data are considered
\cite{gurevich_nonstationary_1974}.  The trailing ($s_-$) and leading
($s_+$) edge DSW velocities from the GP problem are distinct ($s_- < s_+$)
and differ from the single, classical shock velocity derived from the
Rankine-Hugoniot jump conditions for the dispersionless Hopf equation
\begin{equation}
  \label{eq:20}
  u_t + uu_x = 0 .
\end{equation}
Due to its dynamically expanding, distinct edge behavior, a DSW
exhibits an orientation $d$ and polarity $p$, identifying the location
and polarity of the solitary wave edge.  The DSW has $d = +1$ ($d =
-1$) if the solitary wave edge is rightmost (leftmost) and $p = +1$
($p = -1$) if the solitary wave edge is a wave of elevation
(depression) with respect to its adjacent, slowly varying background
(cf.~\cref{fig:kdv_dsw}).  The linear dispersion relation on a
background $\overline{u}$ for \cref{eq:1} is $\omega(k,\overline{u})
= k \overline{u} - \sigma k^3$.  As shown in \cref{fig:kdv_dsw},
the KdV DSW for eq.~\cref{eq:1} has $d = p = -\mathrm{sgn}\,
\omega_{kk} = \mathrm{sgn}\, \sigma$.  We see that two fundamental KdV
DSW properties, its orientation and polarity, are uniquely determined
by the dispersion curvature, also referred to as the sign of
dispersion.  We refer to DSWs that resemble those in
\cref{fig:kdv_dsw} as KdV-like or classical DSWs.

\begin{figure}
\centering
\includegraphics[scale = 0.25]{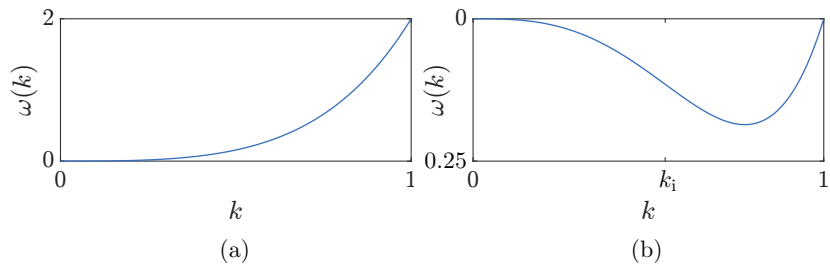}
\caption{Dispersion relation \cref{eq:2} for the Kawahara equation
  \cref{kawahara} for $\bar{u} = 0$, (a) $\sigma = -1$ and (b)
  $\sigma = +1$. }
  \label{fig:kawahara_dispersion}
\end{figure}

In this manuscript, we study DSWs in the presence of higher order
dispersive effects via the fifth order KdV or Kawahara equation
\cite{kawahara}
\begin{equation}
  \label{kawahara} 
  u_t + uu_x + \sigma u_{xxx} + u_{xxxxx} = 0,
\end{equation} 
where $\sigma = \pm 1$.  The dispersion relation on background $\bar{u}$
\begin{equation}
  \label{eq:2}
  \omega(k,\overline{u}) = k \overline{u} - \sigma k^3 + k^5
\end{equation}
is convex when $\sigma = -1$ as depicted in
\cref{fig:kawahara_dispersion}(a).

Purely positive dispersion curvature suggests that Kawahara DSWs occurring in
eq.~\cref{kawahara} with $\sigma = -1$ will be KdV-like and
qualitatively similar to those in \cref{fig:kdv_dsw}(a), which we
indeed find to be the case (see \cref{sec:sigma-=-+1}).  However,
when $\sigma = +1$, the curvature of \cref{eq:2}
\begin{equation}
  \label{eq:3}
  \omega_{kk} = -6k + 20 k^3
\end{equation}
changes sign at the inflection point $k_{\rm i} = \sqrt{3/10}$ as
depicted in \cref{fig:kawahara_dispersion}(b).  Because DSWs are
composed of modulated nonlinear waves with a range of wavenumbers from
zero at the solitary wave edge to a characteristic, nonzero value at
the harmonic edge \cite{el_dispersive_2016}, we expect fundamental
differences in the DSW structure for eq.~\cref{kawahara} when $\sigma
= +1$.  For example, the ``classical'' KdV DSWs described in the last
paragraph feature very different structure (orientation, polarity)
depending on the dispersion curvature.  In this work, we aim to resolve the ways in which a single equation exhibiting both signs of dispersion curvature rectifies these differences. 
%

Note that there is another source of non-convexity in dispersive
hydrodynamic systems: a non-convex, hyperbolic flux.  Such a flux is
known to give rise to undercompressive shock waves and
shock-rarefactions in hyperbolic systems theory and their analogues in
dispersive hydrodynamics \cite{el2015dispersive}.  In contrast, the
problem of non-convex dispersion has no hyperbolic correlate.

In the remainder of this introductory section, we review some relevant
work on DSWs and solitary waves, then provide an overview of this
work.

\subsection{Related work:  dispersive shock waves}
\label{sec:related-work}

Most DSW studies to date have focused upon dispersive hydrodynamic
systems that exhibit either a purely convex or concave linear
dispersion relation \cite{el_dispersive_2016}, with some recent
exceptions
\cite{bakholdin_non-dissipative_2004,trillo,conforti_2013,conforti_2014,conforti_2015,dubrovin_numerical_2011,el2015radiating,malaguti_2014}.
The monograph \cite{bakholdin_non-dissipative_2004} and paper
\cite{dubrovin_numerical_2011} present numerical simulation results
for the Kawahara equation \cref{kawahara} for both short-time
\cite{dubrovin_numerical_2011} and long-time
\cite{bakholdin_non-dissipative_2004} dynamics.  These simulation
results resemble the types of shock waves we characterize in long-time in this work.

In \cite{lowman_dispersive_2013}, a scalar dispersive hydrodynamic
model is shown to exhibit KdV-like DSWs until a critical jump height
is reached corresponding to zero curvature at the harmonic wave edge.
Further increase of the jump height results in the internal,
self-interaction of the DSW or, as it was termed, DSW implosion.  Zero
curvature results in a local extremum of the group velocity that
causes the harmonic edge waves within the DSW to interact with
interior DSW waves of smaller wavenumber.  

Qualitatively different dispersive hydrodynamics near zero dispersion
were observed in Nonlinear Schr\"{o}dinger (NLS) type models of
intense light propagation through fibers
\cite{conforti_2013,trillo,malaguti_2014,conforti_2014,conforti_2015}
and nematic liquid crystals \cite{el2015radiating}. In both cases,
numerical simulations reveal that an essentially linear wavetrain's
phase speed is in resonance with the DSW's solitary wave edge phase
speed, leading to radiation.  As the strength of higher order
dispersion is increased, the DSW structure changes.  Empirical
observations suggest that in long time, the solitary wave edge
exhibits a constant speed moving with the \textit{classical} shock
speed from the Rankine-Hugoniot conditions.  This behavior is in stark
contrast with KdV-like DSWs, whose speeds are determined through
Whitham averaging \cite{el_dispersive_2016}.

\subsection{Related work:  solitary waves}
\label{sec:relat-work:-solit}

Because DSWs can be considered spatially extended generalizations of
solitary waves, it is helpful to briefly review the properties of
solitary wave solutions of eq.~\cref{kawahara}, first computed by
Kawahara \cite{kawahara}.  Distinct structures emerge depending on the
choice of the parameter $\sigma$.  For $\sigma = -1$, the solitary
waves are monotonically decaying from the peak. For $\sigma = +1$, there are
non-monotonically decaying, depression solitary waves for velocities less than
$-\frac{1}{4}$ that are stable \cite{calvo2000stability}.  These
oscillatory solitary waves bifurcate from the linear dispersion curve
\cref{eq:2} when the phase and group velocities 
coincide 
\cite{grimshaw1994solitary}.  The equality of phase and group
velocities occurs only for non-convex dispersion $\omega$. For $\sigma
= +1$ and positive velocities, elevation solitary waves exist but are
unstable due to a linear resonance
\cite{benilov1993generation,pomeau1988structural}.  It is the Kawahara
equation's non-convex dispersion that leads to solitary waves embedded
in the linear spectrum \cite{tan2002semi}.  As we will demonstrate,
non-convex dispersion yields similarly impactful effects on DSW
dynamics.

\subsection{Overview of this work}
\label{sec:overview-this-work}

In \cref{sec:univ-kawah-equat}, we derive the Kawahara equation
\cref{kawahara} from a general dispersive Eulerian system via
multiple-scales perturbation theory as a universal approximate model
for weakly nonlinear dispersive waves when the coefficient of third
order dispersion is small. The requisite conditions for higher order
dispersive effects to be important are identified and the single free
parameter $\sigma$ in eq.~\cref{kawahara} is related to the
dispersive parameters of the original Eulerian system.  We then
consider water waves and nonlinear fiber optics as example dispersive
hydrodynamic systems where this multiple scale method can be applied.

\Cref{sec:solit-wave-solut} reviews the numerical and
asymptotic computation of Kawahara solitary wave solutions and their
corresponding amplitude-speed relations for eq.~\cref{kawahara}.
These solutions are then utilized to help describe the DSWs studied in
\cref{sec:disp-shock-waves}.

In \cref{sec:sigma-=-1-2}, we show that non-convex dispersion
($\sigma = +1$) and sufficiently large jumps lead to a new coherent
structure, a \textit{traveling} DSW (TDSW).  The TDSW is characterized
by a non-monotonic, depression solitary wave trailing edge.  Rather
than complete a full oscillation, the solitary wave is partial and
connects to a periodic nonlinear wavetrain.  This portion of the TDSW
is found to rapidly approach a \textit{genuine traveling wave
  solution} of the Kawahara equation \cref{kawahara}, connecting a
constant to a periodic orbit.  Approximate and numerical periodic
solutions are obtained that yield the TDSW trailing edge speed as a
function of jump height.  The speed is found to be a generalization of
the Rankine-Hugoniot jump condition of classical shock theory.  The
TDSW leading edge is found to move with the linear group velocity.

Small jumps for non-convex dispersion, examined in
\cref{sec:small-jump-heights}, involve long waves and weak fifth
order dispersive effects.  The resulting DSWs are perturbations of
KdV-like DSWs with a leading edge elevation solitary wave that is in
resonance with short, forward-propagating linear waves.  These are
referred to as \textit{radiating} dispersive shock waves (RDSWs).
RDSW properties are determined DSW fitting
theory.

Moderate jumps for non-convex dispersion are examined in
\cref{sec:sigma-=-1-1} where more complex dynamics are observed.
This is the regime that straddles the linear dispersion inflection
point $k_{\rm i}$, corresponding to unsteady, crossover behavior where
we observe strong forward and backward propagation of waves.  We
equate this regime with wave speeds in the solitary wave ``band gap''
where Kawahara solitary waves do not exist but nonlinear periodic
traveling waves do.

Convex dispersion ($\sigma = -1$) is considered in
\cref{sec:sigma-=-+1} where we observe a classical KdV-like DSW.
We apply DSW fitting theory in order to determine the amplitude and
speed of the trailing solitary wave and the wavenumber at the leading
edge as a function of the initial jump height.

Finally, we conclude the manuscript in
\cref{sec:discussionconclusion} with some discussion and broader
perspectives on our findings.

\section{Universality of the Kawahara Equation}
\label{sec:univ-kawah-equat}

We consider a general dispersive Eulerian system of equations given in
non-dimensional form by
\begin{align}
  \rho_t + (\rho u )_x &= D_1[\rho,u]_x,\label{hydro1}\\
  (\rho u)_t +\left(\rho u^2 + P(\rho)\right)_x  &= D_2[\rho,u]_x,\label{hydro2}
\end{align}
where $\rho = \rho(x,t)$ corresponds to the fluid density, $u =
u(x,t)$ the fluid velocity, and the pressure law is given by
$P(\rho)$. We assume strict hyperbolicity $P'(\rho) > 0$ and genuine
nonlinearity $[\rho^2 P'(\rho)]' > 0$ of the dispersionless system
(\cref{hydro1}, \cref{hydro2} with $D_{1,2} = 0$) so that weakly
nonlinear dynamics exhibit quadratic, convex flux
\cite{el2015dispersive}.  The differential operators $D_1$, $D_2$
acting on $\rho,u$ in \cref{hydro1} and \cref{hydro2} are assumed to
be of the second order or higher, yielding a real valued dispersion
relation.  The dispersion is calculated by assuming a small amplitude
linear wave oscillating about the background state $(\rho_0,u_0)$:
$\rho = \rho_0 + A e^{i\theta}$, $u = u_0 + B e^{i\theta}$ where
$\theta = kx - \omega t$ and $|A|$, $|B| \ll 1$ are of the same
order. Substitution of this ansatz into \cref{hydro1} and
\cref{hydro2} yields a homogeneous system of linear equations for $A$
and $B$ that are only solvable for two distinct frequency branches
$\omega_{\pm}(k)$, the dispersion relation. The dispersion relation
exhibits the long wave ($0 < k \ll 1$) behavior
\begin{equation}
  \label{disp_relation}
  \omega_\pm(k)  = u_0 k \pm \Big ( c_0 k + \mu k^3 +
    \gamma k^5 + o(k^5) \Big ),
\end{equation}
where $c_0 = \sqrt{P'(\rho_0)}$ is the long wave speed of sound and
$\mu$, $\gamma$ are the third and fifth order dispersion coefficients,
respectively. In general, these coefficients will depend on $\rho_0$,
$u_0$ and possibly other parameters. We are interested in the
asymptotic balance of third and fifth order dispersion, which can
result when the coefficient of third order dispersion $\mu$ is
sufficiently small.  Since $k$ is inversely proportional to the
characteristic length scale $L$, called the \textit{coherence length}
\cite{el_dispersive_2016}, then fifth order dispersion is important
when $\mu \sim 1/L^2$.  The presence of both third and fifth order
dispersion with comparable magnitudes can result in a change in the
dispersion sign.

We now seek approximate uni-directional solutions to the system
\cref{hydro1} and \cref{hydro2} via multiple-scales in the form
\begin{align*}
  \tau &= \epsilon^{5/4} t, \quad \eta = \epsilon^{1/4}\left(x-(u_0 +
    c_0)t\right), \\
  \rho &=\rho_0 + \epsilon \rho_1(\eta,\tau) +
  \epsilon^{3/2}\rho_2(\eta,\tau) + \epsilon^{2}\rho_3(\eta,\tau) +
  o(\epsilon^2), \\ 
  u & = u_0 + \epsilon u_1(\eta,\tau) + \epsilon^{3/2}u_2(\eta,\tau)
  +\epsilon^{2}u_3(\eta,\tau) + o(\epsilon^2) .
\end{align*}
Note the non-integer powers of $\epsilon$, chosen so that quadratic
nonlinearity will balance the third and fifth order dispersion terms. In other words, we assume the maximal
balance scaling
\begin{equation}
  \label{eq:7}
  \mu = \epsilon^{1/2} \tilde{\mu}, \quad \tilde{\mu} = \mathcal{O}(1) .
\end{equation}
We also assume the boundary conditions
\begin{equation}
  \label{eq:5}
  \rho(x,t) \to \rho_0, \quad u(x,t) \to u_0, \quad x \to \infty .
\end{equation}
Substituting this expansion into \cref{hydro1}, \cref{hydro2} and
using \cref{eq:5}, and applying a standard multiple scales approach as in \cite{whitham2011linear} we have $  \rho_1  = \frac{\rho_0}{c_0}u_1$, where $u_1$ satisfies the Kawahara equation 
\begin{equation}
  \label{kawahara_MS}
  u_{1,\tau} + \alpha u_1 u_{1,\eta} - \tilde{\mu} u_{1,\eta\eta\eta} + \gamma
  u_{1,\eta\eta\eta\eta\eta} = 0 .
\end{equation}
Equation \cref{kawahara_MS} can be put in the normalized form
\cref{kawahara} by use of the scaled variables 
\begin{equation}
  \label{eq:12}
  \begin{split}
    x' = \left|\frac{\tilde{\mu}}{\gamma}\right|^{1/2} \eta, \quad t'
    = \gamma \left|\frac{\tilde{\mu}}{\gamma}\right|^{5/2} \tau,
    \quad u' = \frac{\alpha \gamma}{\tilde{\mu}^2} u ,
  \end{split}
\end{equation}
and then dropping primes.  The key parameter in the Kawahara equation
\cref{kawahara} that encapsulates the competition between third and
fifth order dispersion is
\begin{equation}
  \label{eq:9}
  \sigma = -\sgn(\tilde{\mu} \gamma) .
\end{equation}

We now apply these results to specific model equations from water
waves and fiber optics.  

\subsection{Water waves}
\label{sec:water-waves}

In order to accurately capture the competing effects of third and
fifth order dispersion in water waves, we use the recently derived
extended Green-Naghdi or Serre equations \cite{matsuno2015hamiltonian}
with surface tension effects incorporated as in the generalized Serre (gSerre)
equations \cite{dias2010fully}, yielding the extended, generalized
Serre or egSerre equations.  The corresponding dispersive operators
and general pressure law in \cref{hydro1} and \cref{hydro2} for
egSerre are
\begin{equation}
  \begin{split}
    D_1(\rho,u)& = 0, \\
    D_2(\rho,u) &= \frac{\rho^3}{3}\left(u_{xt} + uu_{xx} -
      u_x^2\right) - B\left[\frac{1}{2} \rho_x^2 - \rho\rho_{xx}\right]
     + \left[ \frac{\rho^5}{45} \left(u_{xxt} + uu_{xxx} - 5
        u_xu_{xx}\right)\right]_x - 3\rho^5u_{xx}^2,\\ 
    P(\rho) &= \frac{\rho^2}{2},
  \end{split}
    \label{eq:10}
\end{equation}
where the dependent variables $\rho(x,t)$ and $u(x,t)$ are the
nondimensional water surface height and vertically averaged horizontal
velocity component, respectively. The bond number, $B$, is a
dimensionless parameter that quantifies the strength of surface
tension relative to gravity. The dispersion relation for
eqs.~\cref{hydro1}, \cref{hydro2} with \cref{eq:10} on the
background $\rho_0 = 1$ and $u_0 = 0$ has the long wave expansion
\begin{equation}
  \label{eq:38}
  \omega(k) = k + \frac 16 (3B - 1) k^3 + \frac{1}{360}(19 - 30 B - 45
B^2) k^5 + o(k^5),
\end{equation}
which agrees with the long wave expansion of the full water wave
dispersion relation \cite{johnson_modern_1997}
\begin{equation}
  \label{eq:13}
  \omega = [(1 + B k^2)k\tanh(k)]^{1/2}.
\end{equation}
The coefficients for the Kawahara equation \cref{kawahara_MS} are
then
\begin{align*}
  \mu &=  \frac{1}{6} (1-3B ), \quad
  \gamma = \frac{1}{360}(19 - 30 B - 45 B^2), \quad \alpha  = \frac 32.
\end{align*}
As noted in our derivation, the Kawahara equation is valid when $\mu$
is small, therefore we are considering $B$ close to $1/3$.  Note that
we have also assumed that $\gamma = \mathcal{O}(1)$.  Since $\gamma$
is zero when $B = (2\sqrt{30}-5)/15 \approx 0.40$, we require $B <
0.4$ for the asymptotic validity of the scaled Kawahara equation
\cref{kawahara} with parameter
\begin{equation}
  \label{eq:11}
  \sigma = -\mathrm{sgn}(\mu \gamma) = \mathrm{sgn}\left ( 1 - 3B
  \right ) . 
\end{equation}
When $B < 1/3$, gravity effects dominate surface tension effects and
we have $\sigma = +1$.  Neglecting the fifth order term, the KdV
equation \cref{eq:1} therefore exhibits negative dispersion with
positive polarity and orientation DSWs as in
\cref{fig:kdv_dsw}(b). 


If we had neglected the higher order terms from the egSerre equations
and just used the gSerre equations, the dispersion would not agree
with the long wave expansion for the full water waves dispersion
relation \cref{eq:13} to fifth order. As such, it is necessary for
one to include \emph{both} the effects that result in small third
order dispersion as well as higher order terms to maintain the
required asymptotic balance.  This suggests that one should consider
with some caution the applicability of the gSerre model to physical
water wave problems when $B$ is near $\frac{1}{3}$.

In \cite{hunter}, the authors derived the Kawahara equation directly
from the Euler equations as a model for shallow water waves for Bond
number near $\frac{1}{3}$.  We have now demonstrated an alternative
derivation based on the egSerre equations via their interpretation as
dispersive Eulerian hydrodynamic equations.

\subsection{Nonlinear fiber optics}
\label{sec:opt}
The effect of higher order dispersive terms in the Nonlinear
Schr\"odinger equation and associated experiments were studied in the
series of papers
\cite{conforti_2013,trillo,malaguti_2014,conforti_2014,conforti_2015}
within the context of light propagation in optical fibers. See also
\cite{bakholdin_non-dissipative_2004} for applications in continuum
mechanics. The equation of interest is a higher order NLS equation
\begin{equation}
  \label{eq:14}
  i \psi_t + \frac{1}{2}\psi_{xx} + i \frac{\beta_3}{6}\psi_{xxx} -
  |\psi|^2 \psi = 0, 
\end{equation}
where $\psi$ is the complex envelope of a weakly nonlinear carrier
wave and $\beta_3$ is a parameter modeling higher order dispersive
effects in the fiber. The variables $x,t$ are used here to maintain
consistency with \cref{hydro1} and \cref{hydro2} but physically
correspond to nondimensionalized time and negative distance along the
fiber, respectively.  The Madelung transformation $\psi =
\sqrt{\rho}e^{i \phi}$, $u = \phi_x$ can be utilized to write
eq.~\cref{eq:14} as a generalized dispersive Eulerian system
\begin{align*}
  \rho_t + \left(\rho u - \frac{1}{2}\beta_3 \rho u^2\right)_x &=
  D_1[\rho,u]_x, \\ 
  \left(\rho u \right)_t + \left(\rho u^2 - \frac{1}{2}\beta_3\rho u^3
    + \frac{1}{2}\rho^2\right)_x & = D_2[\rho,u]_x,  
\end{align*}
where 
\begin{align*}
  D_1[\rho,u] & = \beta_3 \left(\frac{1}{6}\rho_{xx} -
    \frac{1}{8}\frac{\rho_x^2}{\rho}\right),\\ 
  D_2[\rho,u] & =  \frac{1}{4} \rho\left( \log (\rho)\right)_{xx}
  +\frac{\beta_3}{12}\left(\frac{9 \rho_x^2 u}{2 \rho}  - 5 \rho_{xx}
    u - \rho_x u_x - 2 \rho u_{xx}\right). 
\end{align*}
Utilizing the same method as our general derivation of the Kawahara
equation for dispersive Eulerian equations, we obtain
eq.~\cref{kawahara_MS} with coefficients 
\begin{align*}
  \alpha & =  \frac{3 - \beta_3 \left(6u_0 - 3u_0^2\beta_3 +
      \sqrt{\rho_0 - \rho_0u_0 \beta_3}\right)}{2\left(1 -
      u_0\beta_3\right)},\\ 
  \mu & = \frac{(7u_0 \beta_3 -3)\sqrt{\rho_0 - \rho_0u_0 \beta_3}}{24
    \rho_0},\\ 
  \gamma & = \frac{\sqrt{\rho_0 - \rho_0u_0 \beta_3}\left(-9 + 51u_0
      \beta_3 + 16\rho_0\beta_3^2 - 91 u_0^2 \beta_3 ^2 + 49 u_0^3
      \beta_3^3\right)}{1152\rho_0^2\left(u_0\beta_3 -1\right)}. 
\end{align*}
Interestingly, a nonzero background velocity $u_0$ is required in
order to achieve a balance between third and fifth order dispersion.
We note that the numerical simulations in \cite{trillo} consider the
cases $\rho_0 = 0.5$, $u_0 \in (-0.586,-0.543)$, and $\beta_3 \in
(-0.35,-1)$, corresponding to $\sigma = -\mathrm{sgn} \mu \gamma =
+1$, the non-convex dispersion case.

\subsection{Other systems}
\label{sec:other-systems}

The Kawahara equation \cref{kawahara} was derived in the case of
intense light propagation through nematic liquid crystals
\cite{el2015radiating}.  The governing equation is a non-local
NLS-type equation.  The authors numerically observed the generation of
the crossover regime (see \cref{sec:sigma-=-1-1}) and related its qualitative features to those of the
full model equations, where a more detailed numerical and asymptotic
analysis were carried out.

We also note that, with the development of spin-orbit coupled
Bose-Einstein condensates (BECs)
\cite{lin_synthetic_2009,lin_spin-orbit-coupled_2011}, it is possible
to ``engineer'' the dispersion experienced by the wave functions of two nonlinearly coupled spin states.
In a cigar shaped trap where the BEC is approximately one-dimensional,
the mean-field dynamics may be modeled by two coupled NLS equations
(see, e.g., \cite{achilleos_matter-wave_2013} and references therein),
which exhibit non-convex dispersion. 

Another novel application of the Kawahara equation is in the description of shallow water waves under ice cover when the Young modulus of ice is sufficiently large \cite{il2015soliton, marchenko1988long}.

\section{Solitary wave solutions}\label{sec:solit-wave-solut} 

The structure of solitary wave solutions to the Kawahara equation
\cref{kawahara} are well known
\cite{benilov1993generation,grimshaw1994solitary,kawahara}.  In what
follows, we outline relevant properties of these solutions. When
$\sigma = +1$, solitary waves are unstable when embedded in the
continuous spectrum, i.e., when they exhibit a linear resonance for
velocities $c > 0$ \cite{tan2002semi}.  However, there are stable,
non-monotonic solitary waves outside the continuous spectrum when $c <
-\frac{1}{4}$ \cite{calvo2000stability}.  Our investigation of
solitary wave solutions focuses upon their numerical and asymptotic
calculation. A key quantity of interest is the amplitude-speed
relation for these solitary waves, which will prove useful in the
study of DSWs.

We seek solutions of eq.~\cref{kawahara} in the form $u(x,t) =
f(\xi;c)$, $\xi = x- ct$.  Upon integrating once, we obtain
\begin{equation}
  \label{CG_eval}
  -c f + \frac 12 f^2 + \sigma f'' + f^{(4)} = 0,
\end{equation} 
where decay of $f(\xi;c)$ and its first four derivatives eliminates
the integration constant. The nonlinear equation \cref{CG_eval} is
solved via the Newton Conjugate Gradient method \cite{yang2009newton}
or with Matlab's boundary value solver \textrm{bvp5c} for the solitary
wave profile $f$ corresponding to the speed $c$. For comparative
purposes, we recall the well-known soliton solution of the KdV
equation \cref{eq:1}
\begin{equation}
  \label{kdv_soliton}
  f(\xi;c) = a \sigma \sech ^2\left(\sqrt{\frac{|a|}{12}}\xi \right ),
  \quad c(a) = \frac{a \sigma}{3} .
\end{equation}
The monotonic Kawahara solitary waves are well-approximated by the KdV
solution \cref{kdv_soliton} in the small amplitude regime
\cite{benilov1993generation}. Here, monotonic refers to the decay profile on either side of the solitary wave peak--a convention presented in \cite{kawahara}. Higher order dispersion acts as a small
perturbation to the KdV solitary wave.  This effect can be understood
by integrating \cref{CG_eval} again to obtain
\begin{equation}
  \label{eq:26}
  -\frac{c}{2} f^2 + \frac{1}{6} f^3 + \frac{\sigma}{2} \left ( f'
  \right)^2 + f' f''' - \frac{1}{2} \left ( f'' \right )^2 = 0.
\end{equation}
Evaluating at the solitary wave extremum $\xi = 0$ yields the
correction to the KdV speed-amplitude relation \cref{kdv_soliton}
\begin{equation}
  \label{speed correction}
  \begin{split}
    c(a) &= \frac{a\sigma}{3} - \left(\frac{f''(0;c)}{a}\right)^2 = \frac{a\sigma}{3} - \frac{a^2}{36} + \mathcal{O}(a^3) ,
  \end{split}
\end{equation}
The approximate expression is obtained by inserting the KdV soliton
solution \cref{kdv_soliton}. The speed correction in \cref{speed
  correction} is strictly negative, independent of $\sigma$.

\begin{figure}
  \centering
  \includegraphics[scale = 0.25]{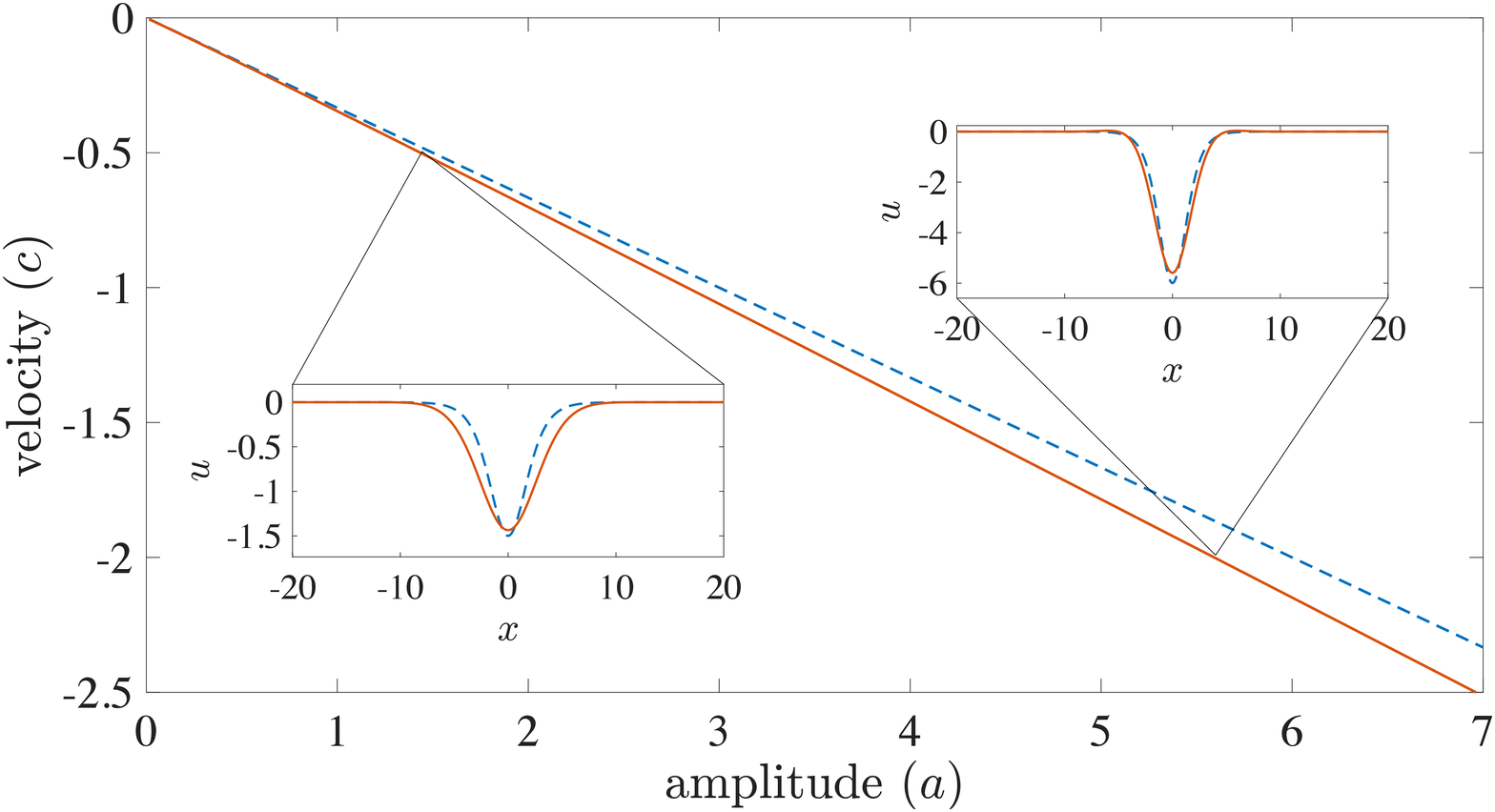}
  \caption{Solitary wave profiles and speed-amplitude relation for the
    Kawahara equation with $\sigma = -1$ (solid) and KdV equation
    (dashed).}\label{speed-amp-pos}
\end{figure}

The Kawahara solitary wave amplitude-speed relation and sample
solitary waves for the case of convex dispersion $\sigma = -1$ are
shown in \cref{speed-amp-pos}.

\begin{figure}
\centering
\includegraphics[scale = 0.25]{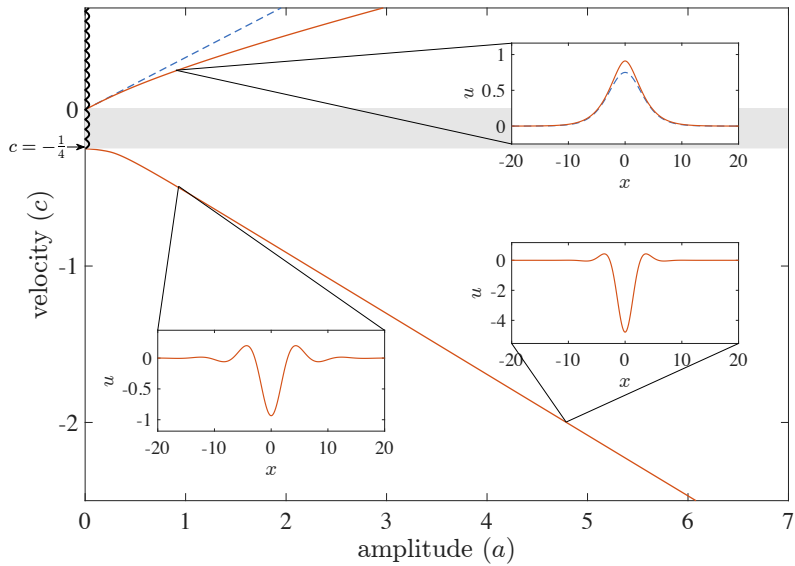}
\caption{Solitary wave profiles and speed-amplitude relation for the
  Kawahara equation with $\sigma = +1$ (solid) and KdV equation
  (dashed). The linear wave spectrum is denoted on the vertical axis
  by the thick black curve.  The ``band gap'' where no solitary waves exist is the shaded region.}
\label{simga_m1_soliton_amp_speed}
\end{figure}

In the regime of non-convex dispersion, $\sigma = +1$, there are two
distinct branches of solitary wave solutions depicted in
\cref{simga_m1_soliton_amp_speed}.  The case $c > 0$ corresponds
to KdV-like elevation waves.  We note that the Kawahara
speed-amplitude relation for $\sigma = +1$ more rapidly departs from
the KdV relation in \cref{simga_m1_soliton_amp_speed} than in
\cref{speed-amp-pos} for $\sigma = -1$.  These solutions are
embedded in the continuous wave spectrum, consisting of all possible
linear phase speeds $\omega/k = -k^2+k^4 > -\frac{1}{4}$ and depicted
on the vertical axis in \cref{simga_m1_soliton_amp_speed}. It
was observed in \cite{benilov1993generation} that a resonance between
the solitary wave and small amplitude waves with the same phase speed
occurs.  This radiation decreases the amplitude of the solitary wave
core as the solution propagates. This is indicative of a general
property of embedded solitary waves \cite{tan2002semi}.

When $c < -\frac{1}{4}$, the solitary waves are depression waves with
non-monotonic profiles. In \cite{calvo2000stability}, it was shown
that this solution branch is stable.  The linearization of the
solitary wave equation \cref{CG_eval} about zero results in a linear,
constant coefficient differential equation with characteristic roots
\begin{equation*}
  r_\pm^2 = \frac{-1\pm \sqrt{1 + 4c}}{2},
\end{equation*}
that are complex with nonzero real part for $c < -\frac 14$ and purely
imaginary for $ - \frac{1}{4} < c < 0$. Consequently, solitary waves with negative velocity 
can only exist for $c < -\frac 14$.  A more detailed, asymptotic
analysis of weakly nonlinear, modulated waves for $0<-\frac 14-c \ll
1$ demonstrates a bifurcation of oscillatory, envelope solitary waves,
hence the non-monotonic profiles in
\cref{simga_m1_soliton_amp_speed} for $c <- \frac 14$
\cite{grimshaw1994solitary,calvo2000stability}.
Our numerical computations yield only nonlinear periodic traveling
waves for velocities $-\frac 14 < c < 0$ so we call this region the
\textit{solitary wave band gap}.

\section{Dispersive Shock Waves}
\label{sec:disp-shock-waves}

In this section, we study DSWs in the Kawahara equation
\cref{kawahara}, fundamental coherent structures in dispersive
hydrodynamics.  Generically, DSWs arise in the long time evolution of
initial data that leads to gradient catastrophe or wavebreaking in the
dispersionless limit.  The canonical GP problem of dispersive
hydrodynamics is to consider the evolution of step initial data
\begin{equation}
  \label{eq:16}
  u(x,0) = \begin{cases} 0,  &x < 0,\\ -\Delta,  &x \geq
    0 \end{cases}, \quad \Delta \in \mathbb{R} .
\end{equation}
More general, two-parameter initial conditions can be obtained by
utilizing the Galilean invariance of the Kawahara equation.  First, we
recall the behavior of solutions to the dispersionless Hopf equation
\cref{eq:20} with initial data \cref{eq:16}.  When $\Delta < 0$, a
rarefaction weak solution
exists and approximates the dispersive hydrodynamics of
eq.~\cref{kawahara} subject to small dispersive corrections.
However, when $\Delta > 0$, the Hopf equation \cref{eq:20} admits a
weak discontinuous shock wave solution
with shock speed
\begin{equation}
 \label{eq:19}
 s = -\frac{1}{2}\Delta
\end{equation}
deduced from the Rankine-Hugoniot (RH) jump conditions.  The
additional dispersive terms in the Kawahara equation act as a singular
perturbation and a different approach must be explored.

In what follows, we use careful numerical computations of the GP
problem for the Kawahara equation \cref{kawahara} with initial data
\cref{eq:16} as one of our analysis tools.  Motivated by the results
of these simulations, we also implement traveling wave computations
and asymptotic analyses that favorably compare with the numerical
results. 

We utilize an integrating factor pseudospectral discretization in
space with 4th order Runge-Kutta (RK4) temporal evolution.  The method
is a generalization of the method applied to the modified KdV equation
in \cite{el2015dispersive} where $u_x$ is assumed to be localized
within the truncated spatial domain so that a Fourier series expansion
is possible.  We temporally evolve $u_x$ using RK4 and the nonlinear
term is computed using a pseudospectral approach.  The numerical
simulations were performed on a spatial domain of length $L=10^5$ and
the location of the initial, $\tanh$ smoothed discontinuity
appropriately chosen to minimize wave-boundary interactions. It is the
fast propagation of small amplitude dispersive waves due to fifth
order dispersion that necessitate such a large domain.  Various
aspects of the approximate solutions were tested in order to ensure
robustness of the numerical method as well as accuracy of the
solution. In particular, all solutions reported exhibit boundary
deviations less than $10^{-3}$.  The Fourier coefficients of $u_x$ all
decay to $10^{-10}$ or less in normalized amplitude and the mass
satisfies the conservation property
$$\int_{0}^L u(x,t) dx - \int_0^L u(x,0) dx = -\frac{t \Delta^2}{2} ,$$
maintained below a relative error of $10^{-3}$, which was only
approached for long times $t = O(100)$ as a result of the small
amplitude oscillations at the boundaries. Typical simulations
presented in this section maintained a relative error on the order of
$10^{-4}$.

\begin{figure}
\centering
\includegraphics[scale=0.25]{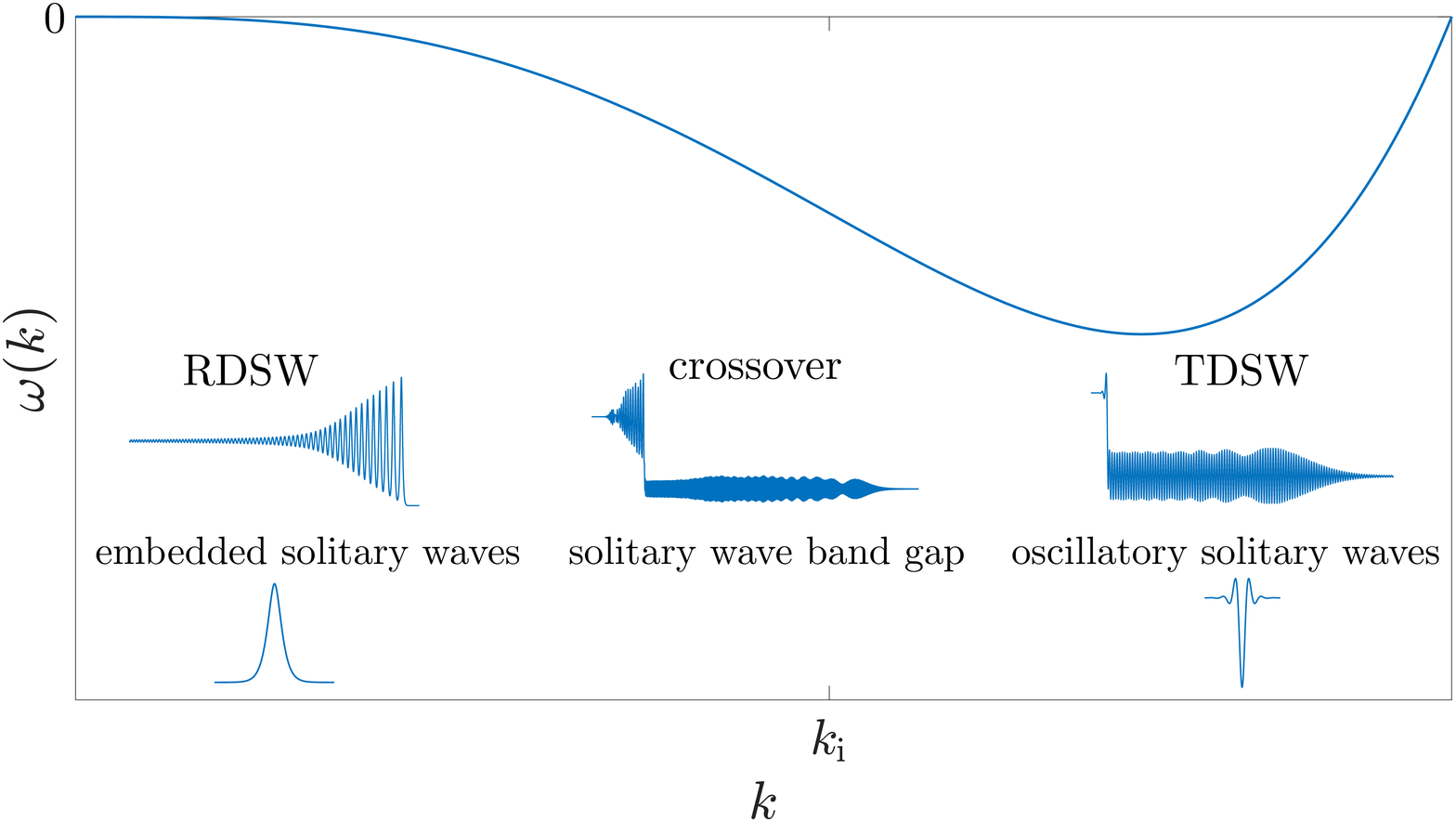}
\caption{Non-convex Kawahara dispersion relation for $\sigma = +1$
  along with example numerical simulations for step initial data.  The
  RDSW, crossover regime, and TDSW can be identified with properties
  of the dispersion curvature and the existence of solitary wave
  structures.}
\label{fig:dispersion}
\end{figure}

We begin our investigation of Kawahara DSWs with the non-convex
dispersion case $\sigma = +1$.  The dynamics can be grouped into three
qualitatively distinct regimes, loosely characterized by the
dispersion relation and soliton solutions.  These regimes are
identified in \cref{fig:dispersion} along with representative
numerical solutions.  Small to large jump heights generate
predominantly small to large wavenumbers, respectively.  The regime of
negative dispersion curvature can be associated with elevation
solitary waves embedded in the linear spectrum, hence naturally appear
as a constituent part of radiating DSWs.  Oscillatory, depression
solitary waves result from non-convex dispersion and are associated
with traveling DSWs.  The crossover regime straddles the dispersion
inflection point and the solitary wave band gap.  

We now undertake a more thorough analysis of the large amplitude
regime and the generation of non-classical traveling DSWs.

\subsection{Non-convex dispersion with large jumps:  TDSWs}
\label{sec:sigma-=-1-2}

\begin{figure}
\centering
\includegraphics[scale = 0.25]{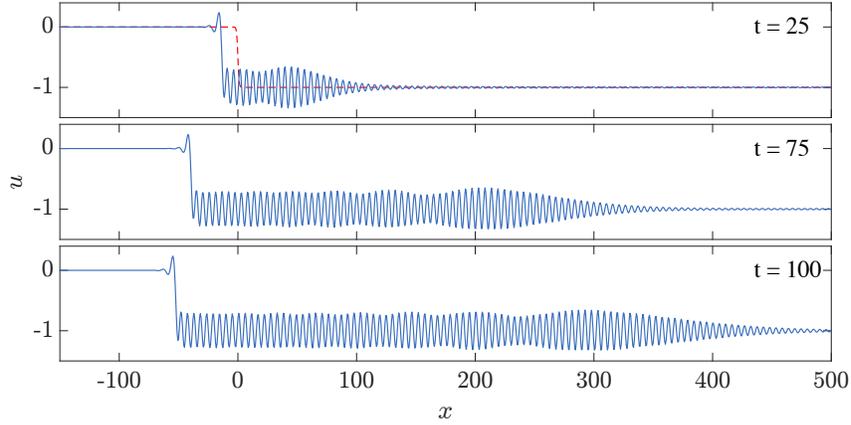}
\caption{Kawahara traveling DSW resulting from the initial step $\Delta
  = 1$ for the non-convex case $\sigma = +1$.  Initial data is shown
  in the top figure with the dashed curve.}
\label{sigma_m1_ul_1}
\end{figure}

We now assume that $\Delta$ is sufficiently large in order to give
rise to a non-classical traveling DSW.  The crossover to this large
jump regime will be made more precise in \cref{sec:sigma-=-1-1}.
\Cref{sigma_m1_ul_1} is a simulation with initial jump $\Delta =
1$.  We observe a sharp, non-monotonic transition from a constant to a
nearly periodic wavetrain.  The wavetrain exhibits some envelope
modulations that eventually terminate at the leading edge with small
amplitude oscillations.  This coherent wavetrain is the TDSW.  We
begin our analysis by verifying two DSW-like properties of the
dynamics: 1) a near solitary wave trailing edge, 2) a harmonic wave
leading edge.

\begin{figure}
\centering
\includegraphics[scale=0.25]{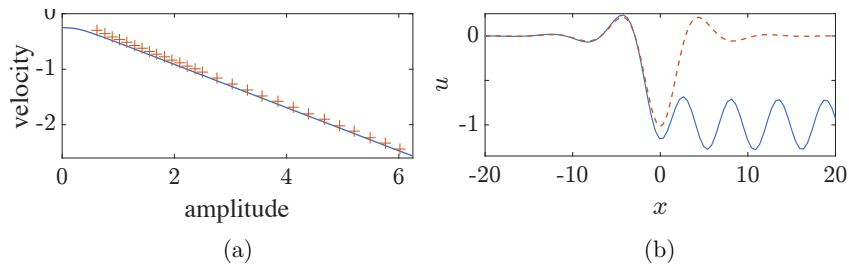}
\caption{Trailing edge comparison with Kawahara ($\sigma = +1$)
  solitary waves.  a) Computed TDSW (pluses) and solitary wave (solid) amplitude-speed relations. b) Overlay of solitary wave on
  TDSW leading edge with $\Delta = 1$.}
\label{sigma_m1_soliton_dsw_comparison}
\end{figure}
First, we plot the amplitude-speed relations for both the trailing
edge of TDSWs with varying $\Delta$, extracted from numerical
simulations, and Kawahara solitary waves in
\cref{sigma_m1_soliton_dsw_comparison}(a), exhibiting excellent
agreement.  Furthermore, in
\cref{sigma_m1_soliton_dsw_comparison}(b), we overlay a Kawahara
depression solitary wave with the same speed as the TDSW trailing
edge.  A portion of the solitary wave correctly captures the
non-monotonic structure of the rapid transition.  Therefore, we can
identify TDSWs with the non-monotonic solitary wave branch of
solutions (cf.~\cref{simga_m1_soliton_amp_speed}). 

\begin{figure}
\centering
\includegraphics[scale=0.25]{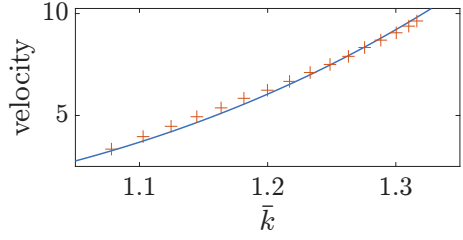}
\caption{Comparison of TDSW harmonic leading edge speed extracted from
  numerical simulations with the linear group velocity. Leading edge
  velocities (shifted by the background $-\Delta$) versus the
  wavenumber $\bar{k}$ near the trailing edge extracted from numerical
  simulations (pluses). The linear group velocity
  $\omega_k(\bar{k},-\Delta)+\Delta$ is also depicted (solid).}
\label{linear_comparison}
\end{figure}

Next, we numerically extract the wavenumber $\bar{k}$ of the TDSW
wavetrain just to the right of the solitary wave trailing edge for
varying $\Delta$ by averaging the wavenumber of 10 oscillations immediately to the right of the partial solitary wave after the TDSW is developed.  The leading edge velocity is also numerically
extracted and compared with the Kawahara group velocity evaluated at
$\bar{k}$ in \cref{linear_comparison}.  We observe excellent
agreement.
Based on these two numerical observations, we see that the TDSW
exhibits solitary and harmonic wave edges, typical of classical DSWs
\cite{el_dispersive_2016}. But that is where the analogy ends.  As we
will now show, the TDSW exhibits unique, non-classical features.

\subsubsection{TDSW trailing edge traveling wave}
\label{sec:tdsw-trailing-edge}

Further scrutiny of the trailing edge shows what appears to be the
development of a nonlinear, periodic wavetrain co-moving with the
partial solitary wave.  The approximately periodic wavetrain
oscillates about the mean value $-\Delta$.  This suggests seeking a
one-parameter family of Kawahara traveling waves $u(x,t) = f(\xi)$,
$\xi = x - ct$ subject to the boundary conditions (BCs)
\begin{align}
  \label{eq:25}
  &\textrm{equilibrium BCs:}\quad \left \{ \lim_{\xi \to -\infty} f(\xi) =
    f'(\xi) = f''(\xi) = f'''(\xi) = f''''(\xi) = 0, \right . \\
  \label{eq:28}
  &\textrm{periodic orbit BC:}\quad 
  \left \{ \begin{aligned}
      \lim_{\xi \to \infty} f(\xi) =
      F(\xi), \quad &F(\xi + P) = F(\xi), \quad \xi \in \mathbb{R}, \\
      &\frac{1}{P} \int_0^P F(\xi) d\xi = -\Delta ,
    \end{aligned} \right .
\end{align}
where $P$ is the wavetrain's period.  Inserting the traveling wave
ansatz into the Kawahara equation \cref{kawahara}, we can integrate
once and apply the boundary conditions at $\xi \to -\infty$ to obtain
the same fourth order ODE \cref{CG_eval} we obtained for solitary
waves.  The traveling wave has two free parameters $c$ and $P$ that
should be uniquely determined by the jump $\Delta$.  One relation is
the mean requirement in \cref{eq:28}.  Another relation is obtained
by integrating \cref{CG_eval} again and applying the boundary
conditions \cref{eq:25} to obtain the \textit{zero energy integral}
\cref{eq:26}.  

To determine the periodic orbit $F(\xi)$, we begin with an
approximate, weakly nonlinear calculation. We consider a small
amplitude $0 < \bar{a} \ll 1$ expansion of $F$ and $c$ as in the classical
Stokes expansion \cite{whitham2011linear}
\begin{align*}
  F(\xi) &=  F_0(\theta) + \bar{a} F_1(\theta) + \bar{a}^2 F_2(\theta) + \cdots,
  \quad \theta = \bar{k}\xi,\\
  c &= c_0 + \bar{a}^2 c_2 + \cdots ,
\end{align*}
where $\bar{k} = 2\pi/P$ is the wavenumber of the periodic orbit.  Inserting
the expansion into \cref{CG_eval} and carrying out a standard
perturbation calculation, we find 
\begin{align}
  \label{eq:35}
  F &= F_0 + \bar{a} \cos \theta + \bar{a}^2 \left(2c_2 -
    \frac{1}{2F_0} -
    \frac{1}{2-16\bar{k}^2+64\bar{k}^4}\cos(2k\xi)\right) +
  o(\bar{a}^2)\\
  \label{eq:36}
  c & = \frac{F_0}{2} + \bar{a}^2 c_2 + o(\bar{a}^2)\\
  \label{eq:15}
  \bar{k}^2 & =\frac{1 + \sqrt{1-2F_0}}{2}
\end{align}
where
\begin{equation}
  \label{eq:6}
  c_2 =  \frac{3 F_0 - 16\bar{k}^2 + 64 \bar{k}^2}{4F_0^2 -
  64\bar{k}^2 + 128\bar{k}^4},
\end{equation}
and $F_0$ is a constant to be determined.
  
The mean requirement \cref{eq:28} applied to \cref{eq:35} yields
\begin{equation} \label{eq:39}
F_0  + a^2\left(2c_2 - \frac{1}{2F_0}\right) = -\Delta. 
\end{equation}
To account for the $\mathcal{O}(\bar{a}^2)$ terms in the mean, the
background $F_0$ is expanded in the parameter $\bar{a}$ in the form
$F_0 = F_{0,0} + \bar{a}^2 F_{0,2} + o(\bar{a}^2)$. Substitution of
the asymptotic expansion yields
\begin{equation}\label{eq:40}
  F_0 = - \Delta - \frac{\bar{a}^2}{2}  \left(\frac{29 \Delta +24
      \sqrt{2 \Delta +1}+24}{\Delta ^2+16 \Delta +8 \sqrt{2 \Delta
        +1}+8}+\frac{1}{\Delta }\right), 
\end{equation}
effectively canceling the $\mathcal{O}(\bar{a}^2)$ mean terms in
\cref{eq:35}.  The only remaining free parameter is the wave
amplitude $\bar{a}$, which can be determined by inserting the
expansion \cref{eq:35} into the zero energy estimate \cref{eq:26}
and evaluating at the wave maximum $\theta = 0$, which yields
\begin{equation}
  \label{eq:41}
  \bar{a} = \frac{\Delta ^{3/2}}{\sqrt{3 + \frac{9 \Delta }{2}+3
      \sqrt{2 \Delta +1}}}. 
\end{equation}
Combining \cref{eq:40}, \cref{eq:6}, \cref{eq:15}. and
\cref{eq:36}, we obtain an amplitude correction to the speed of the
traveling wave
\begin{equation}\label{eq:42}
c = -\frac{\Delta}{2} + \frac{\bar{a}^2}{4\Delta},
\end{equation}
and the square wavenumber of the wavetrain at leading order is given by
\begin{equation}\label{eq:43}
  \bar{k}^2 = \frac{1 + \sqrt{1 + 2 \Delta}}{2}.
\end{equation}

We verify the accuracy of these approximate solutions by directly
computing mean $-\Delta$, zero energy periodic orbits satisfying
eqs.~\cref{CG_eval} and \cref{eq:26} using Matlab's boundary value
solver \textrm{bvp5c}.
\begin{figure}
  \centering
  \includegraphics[scale=0.25]{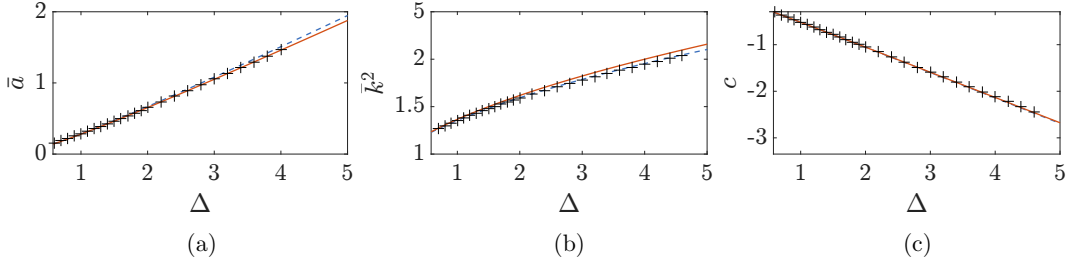}
  \caption{Comparisons of asymptotic predictions of nonlinear
    wavetrain features: (a) the amplitude of the nonlinear wavetrain,
    (b) the square of the wavenumber of the nonlinear wavetrain, and
    (c) the speed of the traveling wave. All figures compare values
    from computed traveling wave solutions (solid),
    asymptotic predictions (dashed), and data extracted
    from numerical simulations of the GP problem (pluses).}
  \label{fig:10}
\end{figure}
\Cref{fig:10} shows the estimated and computed parameters $c$,
$\bar{a}$, and $\bar{k}$ for mean $-\Delta$ periodic solutions.
Excellent agreement is obtained for all values of $\Delta$ for which
the traveling wave solutions were calculated.

What we have shown is that, if a Kawahara traveling wave satisfying
the BCs \cref{eq:25} and \cref{eq:28} exists, its speed $c$ is
determined by the boundary conditions and, in particular, the jump
height $\Delta$.  The first term in the velocity expansion $c_0 =
-\Delta/2$ is the Rankine-Hugoniot jump condition \cref{eq:19} for
classical shock waves.  Therefore, we identify the traveling wave
velocity $c$ as a \textit{generalized Rankine-Hugoniot condition}
(gRH) given in \cref{eq:42}.  

We now directly compute traveling waves satisfying the BCs
\cref{eq:25}, \cref{eq:28} and the zero energy integral
\cref{eq:26}.  Given a jump height $\Delta$ and an associated zero
energy far-field periodic solution $F(\xi)$, we compute solutions of
the fourth order equation \cref{CG_eval} with the four boundary
conditions
\begin{equation}
  \label{eq:44}
  f(0) = f'(0) = 0, \quad f(L) = F(0), \quad f''(L)^2 = -c
    f(L)^2 + \frac{1}{3} f(L)^3 ,
\end{equation}
where $L$ is sufficiently large so that the periodic orbit BC $F(\xi)$
has been reached.  The third condition in \cref{eq:44} evaluates the
periodic orbit at a maximum.  The fourth condition evaluates the zero
energy integral \cref{eq:40} at a maximum.  We use Matlab's
collocation method \textrm{bvp5c} with an initial guess extracted from
the numerical simulation depicted in \cref{sigma_m1_ul_1} that is
then used to perform continuation to other values of $\Delta$.  

\begin{figure}
  \centering
  \includegraphics[scale=0.25]{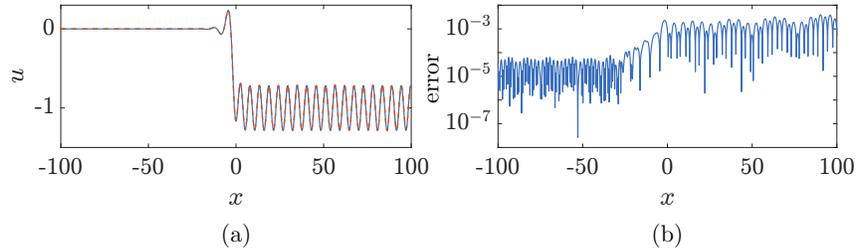}
  \caption{(a) Superimposed traveling wave solution to the dynamical
    system \cref{CG_eval} with boundary conditions \cref{eq:44}
    (solid) on a TDSW computed from the GP problem with
    $\Delta = 1$ (dashed) at $t = 150$. (b) Absolute error
    between the two solutions. }
\label{fig:9}
\end{figure}
The computed traveling wave solution for $\Delta = 1$ is superimposed
on the TDSW determined by long-time integration of the GP problem,
also for $\Delta = 1$, in \cref{fig:9}(a).  Sufficiently near the
TDSW trailing edge, the two solutions are indistinguishable.  
\Cref{fig:9}(b) shows an absolute difference of at most $10^{-3}$
between these two solutions.  This demonstrates that the TDSW trailing
edge rapidly approaches a traveling wave, hence the terminology
\textit{traveling} DSW.

We further examine properties of the TDSW by comparing the numerically
extracted trailing edge velocity, $s_-$, of the TDSW and the speed of
the computed traveling wave, $c$, in
\cref{fig:10}(a). We also compare the amplitude
of the TDSW trailing edge wavetrain to the amplitude of the computed
nonlinear wavetrain in the traveling wave $\bar{a}$ in
\cref{fig:10}(b). Both properties agree over a wide range of $\Delta$.

Our traveling wave computations suggest that the solution is a
heteroclinic connection between the equilibrium $f = 0$ and the mean
$-\Delta$ periodic orbit $f = F(\xi)$.  We are able to accurately
compute such solutions for $\Delta > \Delta_{\rm cr} \approx 0.58$,
suggesting a threshold for their existence.  Such a threshold is
consistent with the speed requirement $c < -\frac14$ for non-monotonic
Kawahara solitary waves (cf.~\cref{simga_m1_soliton_amp_speed}),
of which the TDSW trailing edge is approximately composed.  We will
examine the relationship between $\Delta_{\rm cr}$ and the crossover
to the TDSW regime in \cref{sec:sigma-=-1-1}.

\subsection{Non-convex dispersion with small jumps:  RDSWs}
\label{sec:small-jump-heights}

\begin{figure}
  \centering
  \includegraphics[scale=0.25]{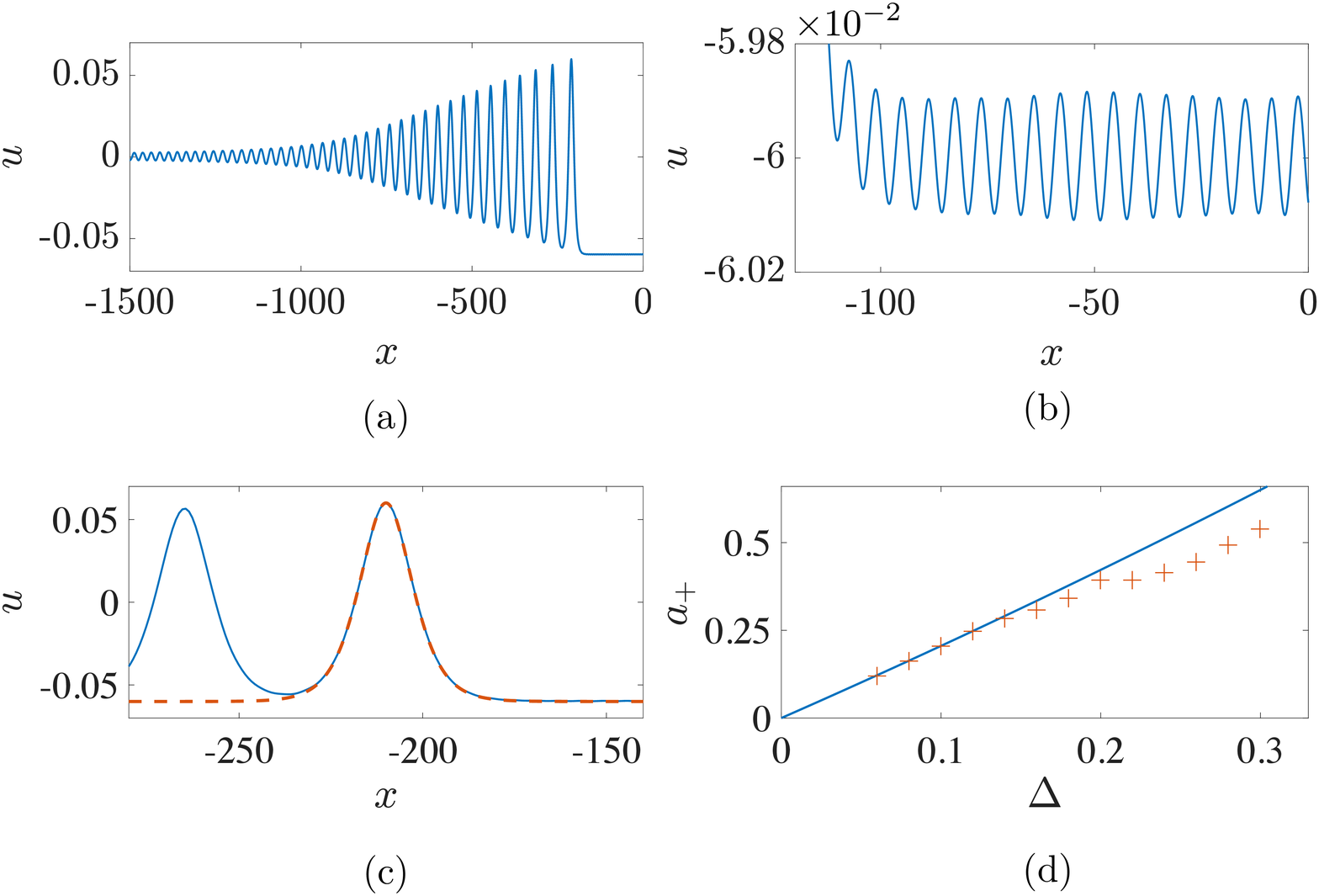}
  \caption{Numerically computed Kawahara radiating DSWs in the small
    jump regime.  (a) The solution at $t \approx 8573$ for jump 
    $\Delta = 0.06$. The radiation is not visible due to its small
    amplitude. (b) Zoomed in view of radiation for RDSW in panel (a).
    (c) RDSW leading edge from (a) with an overlay of a numerically
    computed Kawahara solitary wave of the same speed. (d) Comparison
    of the leading edge ampltude $a_+$ versus $\Delta$ incorporating
    predictions from Kawahara DSW
    fitting (solid) and extracted values from numerical
    simulations (pluses).}
\label{epsilon_dsw}
\end{figure}

We now consider the non-convex case of eq.~\cref{kawahara} ($\sigma =
+1$) in the small jump regime, $0 < \Delta \ll 1$.  Introducing the
scaling
\begin{equation}
  \label{eq:21}
  u = \Delta U, \quad X = \Delta^{1/2} x, \quad T = \Delta^{3/2} t,
\end{equation}
into eq.~\cref{kawahara} results in a perturbed KdV equation
\begin{equation}
  \label{KdV_neg_dispersion}
  U_T + UU_X + U_{XXX} = -\Delta U_{XXXXX}.
\end{equation}
In the scaled variables \cref{eq:21}, the initial conditions
\cref{eq:16} become
\begin{equation}
  \label{eq:22}
  U(X,0) =
  \begin{cases}
    0 & X < 0, \\
    -1 & X > 0
  \end{cases}.
\end{equation}
Numerically, we evolve the scaled equation \cref{KdV_neg_dispersion}
subject to \cref{eq:22} but report the results for the unscaled field
$u(x,t)$ through \cref{eq:21}. Numerical results are shown in
\cref{epsilon_dsw}.  Sufficiently small jumps lead to KdV-like,
classical DSWs as \cref{epsilon_dsw}(a) with $\Delta = 0.06$
attests.  In this long-wave regime, the Kawahara linear dispersion
relation \cref{eq:2} is essentially concave
(cf.~\cref{fig:kawahara_dispersion}) so that the resulting DSWs exhibit
positive polarity and orientation.  The DSW leading edge is well
approximated by an elevation Kawahara solitary wave as shown in
\cref{epsilon_dsw}(c). However, due to the embedding of the elevation solitary waves in the continuous spectrum, the solitary wave emits small amplitude radiation ahead of the shock, a phenomenon demonstrated in \cref{epsilon_dsw}(b). This DSW resonant radiation has also been observed in NLS-type models \cite{trillo,el2015radiating},
so we introduce the nomenclature \textit{radiating} DSW (RDSW) to
describe this phenomenon.

For the analysis of RDSWs, one could consider Whitham theory
\cite{whitham2011linear} for the full Kawahara equation.  We directly
apply El's DSW fitting method \cite{el2005resolution} (see also
\cite{el_dispersive_2016}), which assumes the applicability of Whitham
theory.  Under appropriate conditions, the method yields the leading
and trailing edge speeds as functions of the jump $\Delta$.
Additional macroscopic DSW properties that can be obtained are the
solitary wave edge amplitude and the harmonic wave edge characteristic
wavenumber. The fitting method can be carried out with knowledge of
only the linear dispersion relation and the solitary wave
amplitude-speed relation, both of which we know exactly or
approximately.  We note that the underlying assumptions for the
validity of the DSW fitting method require additional considerations,
which we do not fully explore here.  Rather, we apply the method and
compare the results with our numerical simulations.

The RDSW trailing edge wavenumber $k_-$ is characterized by a simple
wave solution of the Whitham modulation equations.  This wavenumber
can be determined from the solution of the ODE
\begin{equation}
  \label{whitham_ode}
  \frac{dk}{d \bar{u}} = \frac{\omega_{\bar{u}}}{\bar{u} - \omega_k} =
  \frac{1}{3k - 5k^3},
  \quad k(-\Delta) = 0, 
\end{equation}
where $\omega$ is the Kawahara linear dispersion relation
\cref{eq:2}.  The modulation variable $\bar{u}$ corresponds to the
period-mean of the modulated periodic traveling wave and the boundary
condition $k(-\Delta) = 0$ is due to the vanishing of the modulation
wavenumber at the RDSW solitary wave edge where $\bar{u} = -\Delta$.
The ODE \cref{whitham_ode} can be directly integrated, yielding
\begin{equation}
  \label{eq:29}
  k(\bar{u})^2 = \frac{3 - \sqrt{9 - 20 (\bar{u} + \Delta)}}{5}.
\end{equation}
The RDSW trailing edge wavenumber is determined by evaluating
\cref{eq:29} at the RDSW trailing edge where $\bar{u} = 0$
\begin{equation}
  \label{eq:30}
  \begin{split}
    k_- &= k(0) = \left ( \frac{3 - \sqrt{9-20\Delta}}{5} \right
    )^{1/2} 
    = \left (\frac{2\Delta}{3} \right )^{1/2} + \frac{5
      \Delta^{3/2}}{9\sqrt{6}} + \mathcal{O}(\Delta^{5/2}) .
  \end{split}
\end{equation}
Reflecting the harmonic wave nature of the trailing edge, its velocity
$s_-$ is then determined by evaluating the Kawahara linear group
velocity at the trailing edge
\begin{equation}
  \label{eq:31}
  \begin{split}
    s_- &= \frac{\partial \omega}{\partial k}(k_-,0) = -2\Delta +
    \frac{(3- \sqrt{9 - 20 \Delta})^2}{10} 
      = -2\Delta + \frac{10}{9} \Delta^2 + \mathcal{O}(\Delta^3).
  \end{split}
\end{equation}

There are several ``barriers'' to the DSW fitting analysis
\cite{el_dispersive_2016}.  The first barrier occurs at an extremum of
the trailing edge speed as a function of jump height.  The minimum of
$s_-(\Delta)$ occurs when $\Delta = \Delta_{\rm i} = 27/80 \approx
0.34$.  At this value of $\Delta$, the trailing edge wavenumber $k_- =
\sqrt{3/10}$ is precisely the zero dispersion point $k_{\rm i}$.  We
cannot expect the DSW fitting method to accurately describe RDSWs for
$\Delta > \Delta_{\rm i}$.  In another model equation with non-convex
dispersion, crossing this barrier led to DSW implosion
\cite{lowman_dispersive_2013}.  Note that the jump height $\Delta =
9/20$, above which $k_-$ becomes complex-valued, exceeds the barrier
$\Delta_{\rm i}$.

The second barrier occurs when the hyperbolic Whitham modulation
system loses genuine nonlinearity at a linearly degenerate point.
This barrier can be identified at the RDSW harmonic wave edge by
finding the zero of $\omega_{k\bar{u}} (\bar{u} - \omega_k) +
\omega_{kk} \omega_{\bar{u}}$ \cite{el_dispersive_2016}, which occurs
at the jump height $\Delta = \Delta_{\rm l} = 1/4$.  This second
barrier occurs at a smaller jump height than the first.

The speed at the DSW leading edge is calculated in a similar manner by
first introducing conjugate variables $\tilde{k}$, and
$\tilde{\omega}(\tilde{k},\bar{u}) = - i \omega(i\tilde{k},\bar{u})$,
where $\tilde{k}$ acts as an amplitude parameter and $\tilde{\omega}$
is a ``solitary wave dispersion relation''. One now solves the ODE
\begin{equation}
  \label{eq:33}
  \frac{d \tilde{k}}{d\bar{u}} =
  \frac{\tilde{\omega}_{\bar{u}}}{\bar{u}-\tilde{\omega}_{\tilde{k}}}
  = -\frac{1}{3\tilde{k} + 5\tilde{k}^3}
  , \quad \tilde{k}(0) = 0.
\end{equation}
Integrating and evaluating the conjugate wavenumber at the solitary
wave leading edge yields $\tilde{k}_+^2 = \tilde{k}(-\Delta)^2 =
\frac{-3 + \sqrt{9 + 20 \Delta}}{5}$.  The DSW leading edge speed
$s_+$ is the conjugate phase velocity evaluated at the leading edge
\begin{equation}
  \label{eq:34}
  \begin{split}
    s_+ &= \frac{\tilde{\omega} (\tilde{k}_+,-\Delta)}{\tilde{k}_-} 
    = \frac{3}{25} - \frac{1}{5}\Delta - \frac{1}{25}\sqrt{9 +
      20\Delta}  \\
    &= -\frac{1}{3}\Delta + \frac{2}{27} \Delta^2 +
    \mathcal{O}(\Delta^3) .
  \end{split}
\end{equation}
Utilizing the approximate Kawahara solitary wave amplitude-speed
relation \cref{speed correction}, we can obtain an estimate for the
solitary wave edge amplitude $a_+$ by equating $s_+ = c(a_+)$,
yielding
\begin{equation}
  \label{eq:24}
  a_+ = 2 \Delta + \frac{5}{9} \Delta^2 +
  \mathcal{O}(\Delta^3) . 
\end{equation}

We note that all of the small $\Delta$ asymptotics in
eqs.~\cref{eq:30}, \cref{eq:31}, \cref{eq:34}, and \cref{eq:24} of
the RDSW agree with the results for KdV at leading order in $\Delta$
\cite{gurevich_nonstationary_1974}.  However, the RDSW exhibits an
additional, radiative component.  Using the RDSW analysis, we can
estimate some of the properties of the forward, short-wave radiation.
The resonance condition
\begin{equation}
  \label{eq:54}
  s_+ = \frac{\omega(k_{\rm r},-\Delta)}{k_{\rm r}} = -\Delta - k_{\rm
    r}^2 + k_{\rm r}^4,
\end{equation}
equates the RDSW leading edge solitary wave speed \cref{eq:54} with
the phase speed of linear waves, thus determining the resonant
wavenumber $k_{\rm r}$
\begin{equation}
  \label{eq:55}
  \begin{split}
    k_{\rm r}^2 &= \frac{1}{2} + \frac{1}{10} \left (37 + 80 \Delta -
      4\sqrt{9 + 20 \Delta} \right )^{1/2} \\
    &= 1 + \frac{2}{3} \Delta + \mathcal{O}(\Delta^2) .
  \end{split}
\end{equation}
Because $k_{\rm r}$ exceeds the linear dispersion inflection point
$k_{\rm i} = \sqrt{3/10}$, resonant radiation corresponds to the
positive dispersion regime.  We note that this resonant radiation
wavenumber agrees with the wavenumber $\bar{k}$ associated with the
TDSW in eq.~\cref{eq:43} only at leading order ($\bar{k} \sim 1 +
\frac14 \Delta$, $0 < \Delta \ll 1$).

\begin{figure}
  \centering
  \includegraphics[scale=0.25]{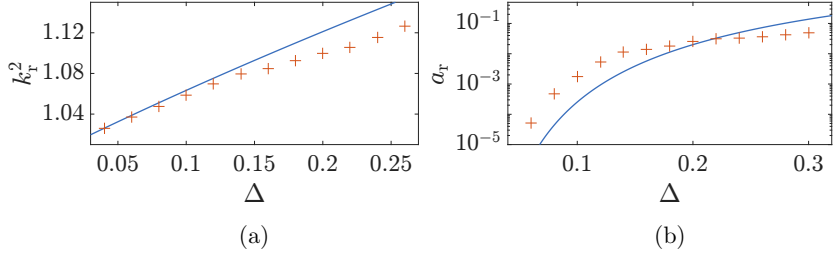}
  \caption{Comparison of amplitude of the radiation in the small
    amplitude RDSW with Analytical prediction from \cref{eq:23}
    (solid) and extracted values from numerical simulations
    (pluses).}
  \label{radiation_amp}
\end{figure}

References \cite{pomeau1988structural,benilov1993generation} provide
an asymptotic estimate for the amplitude $a_{\rm r}$ of the radiation
from an unstable Kawahara solitary wave.  Because a lone solitary wave
decays due to this linear resonance, the amplitude $a_{\rm r}$ was
found to be a time-dependent quantity.  However, the RDSW leading edge
approximate solitary wave is sustained so, using the results of
\cite{pomeau1988structural,benilov1993generation}, we estimate the
\emph{constant} radiation amplitude
\begin{equation}
  \label{eq:23}
  a_{\rm r} \sim K \exp \left ( -\frac{3\pi}{2\sqrt{\Delta}} \right ) 
  , 
\end{equation}
where $K \approx 752.85$ is a numerical constant.  
\Cref{radiation_amp} compares the numerically extracted RDSW radiation
wavenumber $k_{\rm r}$ and amplitude $a_{\rm r}$ with the predictions
of Eqs.~\cref{eq:55} and \cref{eq:23}.  We note that due to fast dispersive propagation to the boundary, it becomes
exceedingly difficult to numerically resolve the exponentially small
radiation amplitude $a_{\rm r}$ for small $\Delta$, likely the cause
of the discrepancy in \cref{radiation_amp}(b).  This shows, and
has been noted previously \cite{el2015radiating}, that a RDSW provides
a means to effectively sustain a Kawahara solitary wave--which would
otherwise decay due to linear resonance
\cite{benilov1993generation}--as part of a DSW.
As \cref{epsilon_dsw}(d) reveals, the RDSW leading edge amplitude
closely follows the prediction in eq.~\cref{eq:24} until the jump
exceeds about 0.2.  One possible explanation for this could be the
apparent loss of genuine nonlinearity in the Whitham equations when
$\Delta > \Delta_{\rm l} = 1/4$.  At larger jumps, the RDSW solitary
wave edge no longer resembles a Kawahara solitary elevation wave
solution, but begins to share qualities with TDSWs.  The dynamics begin to lose
the rank ordered structure that is characteristic of classical DSWs.
We now analyze the intermediate, crossover regime where the DSW
structure gradually transitions from RDSWs to TDSWs.

\subsection{Non-convex dispersion with intermediate jumps:  the
  crossover regime}
\label{sec:sigma-=-1-1}

\begin{figure}
\centering
  \includegraphics[scale = 0.25]{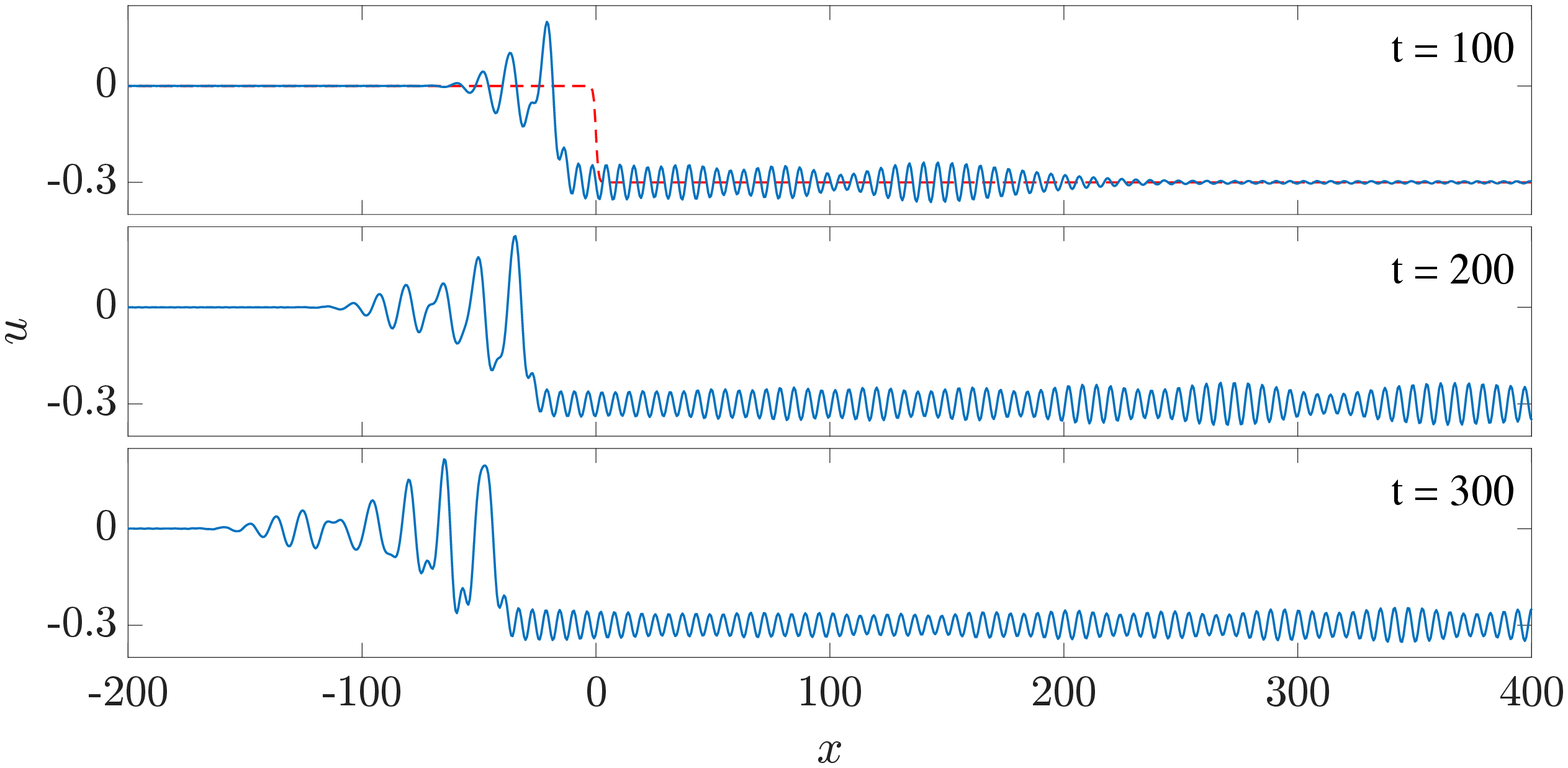}
  \caption{Crossover Kawahara DSW dynamics for the intermediate jump
    value $\Delta = 0.3$ with $\sigma = +1$. Initial data is the
    dashed curve. }
  \label{sigma_m1_ul_03} 
\end{figure}

As the magnitude of the jump initial data increases, the RDSW begins
to lose KdV-like characteristics while gaining features of a TDSW. This transition between the RDSW with positive polarity and orientation
(small jumps) and the TDSW with negative polarity and orientation
(large jumps) occurs gradually as $\Delta$ is increased in
magnitude. The evolution of the GP problem in \cref{sigma_m1_ul_03} is
representative of the evolution of step initial data with $\Delta = 0.3$ and displays significant
backward radiation adjacent to a recessed, large amplitude,
oscillatory region.  The structure of the oscillatory region exhibits
slower amplitude decay and more of an amplitude separation from the
leading edge than that of the smaller jump depicted in
\cref{epsilon_dsw}(a).  The DSWs for values of $\Delta$ in this
region are qualitatively characterized by this remnant of a small
amplitude RDSW with positive polarity and orientation with
superimposed small amplitude waves that suggest incoherence. Such incoherence 
results in wave mixing that eliminates the rank-ordered
structure of the RDSWs that occur at smaller jumps.  The largest amplitude, elevation wave does
not appear to resemble any of the Kawahara solitary waves we have
computed in \cref{simga_m1_soliton_amp_speed} and waves propagate
both ahead of and behind the peak.  Just as we identify the leading
edge of RDSWs resulting from small jumps with elevation solitary waves
in \cref{simga_m1_soliton_amp_speed} ($c > 0$) and the TDSW
trailing edge for large jumps with non-montonic elevation solitary
waves ($c < -1/4$), we interpret the intermediate jump transition
region as corresponding to the solitary wave ``band gap'' for $-1/4 <
c < 0$ in \cref{simga_m1_soliton_amp_speed}.  For velocities in
the band gap, solitary waves do not exist.  However, we can compute
periodic traveling waves in this region so a modulation description
may be possible, but we do not pursue this further here.  Rather, we
seek to identify when the backward radiation on $u = 0$ emanating from
the transition to $u = -\Delta$ ceases, signifying the onset of the
steady TDSW.

If these backward radiating waves were, in fact, linear then they could
persist whenever the linear phase velocity on zero background
coincides with the edge speed, otherwise we expect a TDSW.  The linear
phase velocity $v_{\rm ph} = \omega(k,0)/k = -k^2 + k^4$ attains a
minimum value of $-1/4$ precisely when the phase and group velocities
coincide $k = 1/\sqrt{2}$ and when the non-monotonic Kawahara
solitary waves appear (cf.~\cref{simga_m1_soliton_amp_speed}).
Equating the minimum of $v_{\rm ph}$ to the TDSW leading order gRH  \eqref{eq:42},  $c =
-\Delta/2$ gives the critical jump height $ \Delta_{\rm cr} \sim \frac{1}{2},$
above which linear waves cannot propagate behind the TDSW.  However,
the numerical simulations show that waves continue to propagate
backward even when $\Delta = 1/2$.  Our numerical simulations have
shown that this phenomenon persists up to jumps of $\Delta \approx
0.6$.  Although close to the theoretical prediction, we
argue that the true threshold criterion is the existence of the TDSW
traveling wave.  We found in \cref{sec:tdsw-trailing-edge} that we
could no longer compute TDSWs for $\Delta$ below 0.58, very close to
the observed transition to TDSWs at $\Delta = 0.6$.  Therefore, the TDSW is a
\textit{threshold} phenomenon, only existing for $\Delta > \Delta_{\rm
  cr}$.  For $\Delta < \Delta_{\rm cr}$, either perturbed, classical
DSWs are generated as in \cref{sec:small-jump-heights} (when
$\Delta \lesssim 0.2$ from \cref{epsilon_dsw}(d))
or a crossover, oscillatory state lacking a well-defined solitary wave
edge.

\subsection{Convex dispersion}
\label{sec:sigma-=-+1}

In the case where the sign of the third order term is negative
($\sigma = -1$), the Kawahara dispersion relation \cref{eq:2} is a
purely convex function of $k$.  These are ``convex dispersive
hydrodynamics'' so we expect KdV-like DSWs.  For completeness, we
briefly analyze this case.

\begin{figure}
  \centering
  \includegraphics[scale = 0.25]{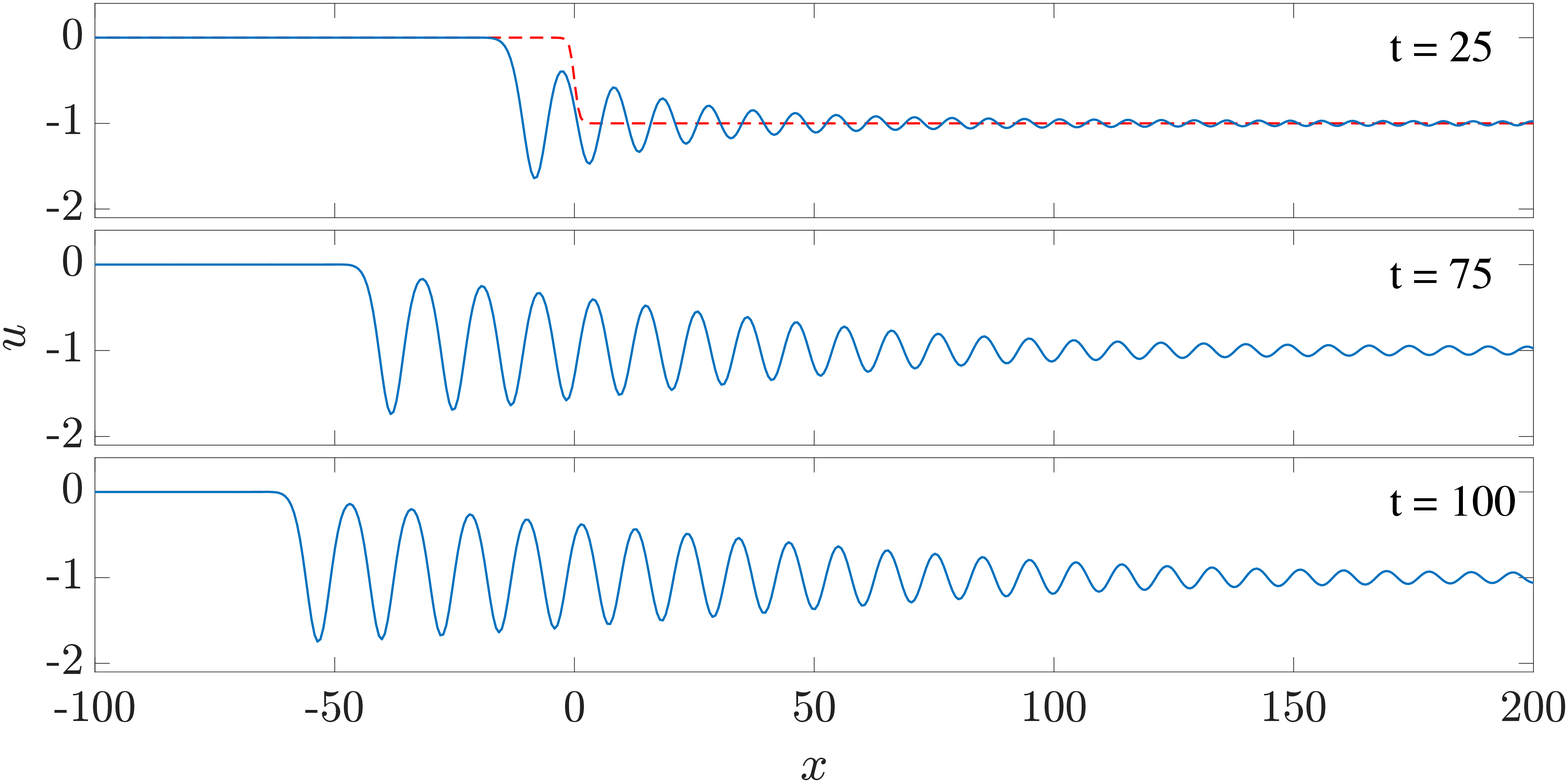}
  \caption{Development of a classical Kawahara DSW with initial jump
    $\Delta = 1$ and convex dispersion $\sigma = -1$. Approximate
    initial data is shown in the top figure with the dashed
    curve.}
  \label{sigma_p1_ul_1}
\end{figure}
\begin{figure}
\centering
\includegraphics[scale=0.25]{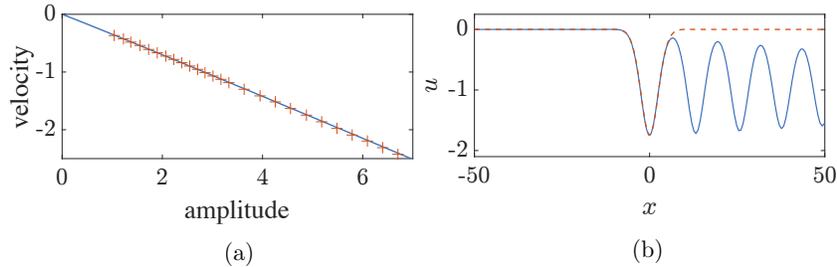}
\caption{Comparison of DSW leading edge properties to Kawahara
  solitary waves. (a) Speed-amplitude relation of Kawahara solitary
  wave (solid) and DSW trailing edge (pluses) for $\sigma = +1$. (b)
  Overlay of numerically computed solitary wave with coincident
  velocity with the DSW trailing edge for $\Delta= 1$.}
  \label{leading_edge_convex}
\end{figure}

The numerical simulation in \cref{sigma_p1_ul_1} depicts the
temporal evolution of step initial data \cref{eq:16} for
eq.~\cref{kawahara}. This figure portrays the temporal development of
a DSW that is qualitatively similar to the classical KdV DSW. The
addition of the fifth order term in eq.~\cref{kawahara} serves as a
perturbation to the KdV equation that, in contrast to the non-convex
case $\sigma = +1$, does not qualitatively change the dynamics.  The
DSW trailing edge behaves like a solitary wave solution of the
Kawahara equation as shown by \cref{leading_edge_convex}(a) where
the DSW trailing edge speed-amplitude relation, extracted from
multiple simulations, is compared to the solitary wave amplitude speed
relation from \cref{speed-amp-pos}.  A Kawahara solitary wave of
velocity given by the trailing edge is superimposed on the DSW
trailing edge in \cref{leading_edge_convex}(b).


\begin{figure}
\centering
  \includegraphics[scale=0.25]{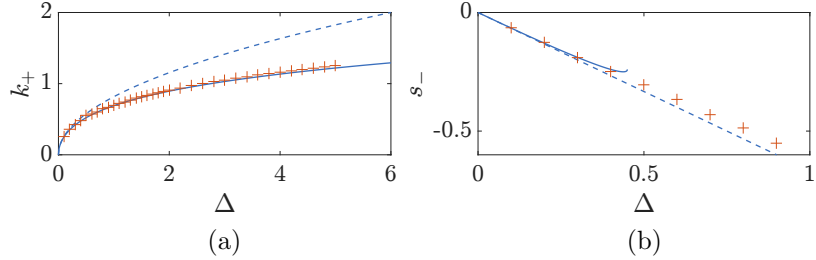}
  \caption{Kawahara DSW trailing edge wavenumber (a) and DSW leading
    edge speed (b) for varying jump height.  Comparison between
    Whitham theory predictions for the Kawahara equation (solid), Whitham theory for the KdV equation (dashed) and numerical
    simulation (pluses).}
  \label{whitham}
\end{figure}

We now implement the DSW fitting method \cite{el2005resolution} (see
also \cite{el_dispersive_2016}).  The implementation is essentially
the same as that of \cref{sec:small-jump-heights} but we now use
the dispersion relation \cref{eq:2} and approximate solitary wave
amplitude-speed relation \cref{speed correction} both with $\sigma =
-1$.  We omit the details.  The macroscopic DSW harmonic leading edge
properties are the characteristic wavenumber
\begin{equation}
  \label{eq:58}
  \begin{split}
    k_+ &= \left ( \frac{-3 + \sqrt{9+20\Delta}}{5} \right
    )^{1/2} 
    = \left (\frac{2\Delta}{3} \right )^{1/2} - \frac{5
      \Delta^{3/2}}{9\sqrt{6}} + \mathcal{O}(\Delta^{5/2}) ,
  \end{split}
\end{equation}
and speed
\begin{equation}
  \label{eq:59}
  \begin{split}
    s_+ &= \frac{9}{5} + 3 \Delta - \frac{3}{5} \sqrt{9 + 20 \Delta} 
    = \Delta + \frac{10}{9} \Delta^2 + \mathcal{O}(\Delta^3).
  \end{split}
\end{equation}
\Cref{whitham}(a) shows the DSW harmonic edge wavenumber $k_+$
versus jump height.  DSW fitting theory provides an excellent
approximation of the trailing edge wavenumber, extracted from our
numerical simulations.  In particular, DSW fitting correctly captures
the reduction of the trailing edge wavenumber relative to the leading
order KdV result $k_+= \sqrt{2\Delta/3}$.  We see that higher order
dispersion has a significant quantitative effect on the properties of
the harmonic wave edge.

The macroscopic properties of the DSW solitary wave trailing edge
include the velocity
\begin{equation}
  \label{s_plus}
  \begin{split}
    s_- &= \frac{3}{25} - \frac{4}{5}\Delta - \frac{1}{25}\sqrt{9 -
      20\Delta}  
    = -\frac{2}{3}\Delta + \frac{2}{27} \Delta^2 +
    \mathcal{O}(\Delta^3) ,
  \end{split}
\end{equation}
and the amplitude
\begin{equation}
  \label{eq:27}
  a_- = 2 \Delta - \frac{5}{9} \Delta^2 + \mathcal{O}(\Delta^3) ,
\end{equation}
approximated by using eq.~\cref{speed correction} and equating $s_- =
c(a_-)$.  Although the trailing edge velocity is only defined for $0 <
\Delta < 9/20$, the small $\Delta$ asymptotics agree with the KdV
velocity (and amplitude $a_+$) to leading order
\cite{gurevich_nonstationary_1974}.  The next order correction shows
that the Kawahara DSW solitary wave edge velocity is above the
corresponding KdV DSW velocity, which agrees with the numerical simulations
shown in \cref{whitham}(b) for $\Delta$ below the critical value
$9/20$.  The DSW fitting method fails for $\Delta > 9/20$, even though
numerical computations show a clear trend.

\section{Discussion/Conclusion}
\label{sec:discussionconclusion}

The Kawahara equation is a universal asymptotic model of weakly
nonlinear, dispersive hydrodynamics with higher order dispersion.  The
classification of the Gurevich-Pitaevskii initial step problem carried
out here reveals classical, KdV-like DSWs when the dispersion is
convex and three distinct regimes for non-convex dispersion.  These
three regimes represent an intrinsic mechanism for the transition from
convex to non-convex dispersive hydrodynamics.  An example from
shallow water waves (recall \cref{sec:water-waves}) illuminates
this transition.

When gravity dominates surface tension effects, the Bond number $B$ is
small so that higher order dispersive effects continue to yield
negative dispersion curvature for all but very short wavelengths
(recall eq.~\cref{eq:13}).  DSWs in this regime are therefore
KdV-like, satisfying eq.~\cref{eq:1} with $\sigma = +1$, with 
positive orientation and polarity as in \cref{fig:kdv_dsw}(b).
For $B$ less than but close to $1/3$, where surface tension and
gravity start to balance, the non-convexity of the dispersion relation
manifests in the Kawahara equation \cref{kawahara} with $\sigma =
+1$.  Small jumps still yield KdV-like DSWs with positive orientation
and polarity but now they are accompanied by a resonance and small
amplitude forward radiation.  They are RDSWs.  As the jump height is
further increased, the forward radiation gets stronger at the expense
of backward wave propagation until a critical jump height is reached.
Above this threshold, a TDSW with negative orientation and polarity is
generated, exhibiting a steady traveling wave structure at the
trailing edge.  Thus, the crossover from positive to negative DSW
polarity and orientation manifests as an intrinsic feature of the
Kawahara equation as the jump height is increased.  For $B > 1/3$,
$\sigma = -1$ and the DSWs are all KdV-like with negative polarity and
orientation.  Because the Bond number is inversely proportional to
fluid depth, we expect to see these non-convex features for
sufficiently shallow flows.

Higher order dispersive effects can play an important role in nonlinear fiber optics as demonstrated in \cite{trillo,conforti_2013, conforti_2014,conforti_2015, malaguti_2014}.  The Kawahara equation is a simpler, scalar, unidirectional model in which to interpret the dynamics of these works (cf. \cref{sec:opt}).  In particular, the authors in \cite{trillo} observed coherent structures consistent with RDSWs and TDSWs described in detail here.  Solitary wave and DSW resonances modeled by a third order NLS equation were observed experimentally in [9].  This motivates further analysis of higher order NLS models.  Can TDSW traveling wave solutions of the bidirectional NLS equation with third order dispersion be obtained?  Also, the implications of these coherent structures for physical applications warrants further exploration.

The TDSW is a non-classical DSW in the sense that it is not KdV-like,
rather it satisfies a generalized Rankine-Hugoniot relation resulting
from the far-field behavior of a constant in one direction and a
periodic traveling wave in the other.  The TDSW rapidly approaches a
traveling wave solution of the Kawahara equation satisfying these
far-field conditions, consisting of a coherent combination of a
uniform wavetrain connected to the constant value through a partial,
non-monotonic solitary wave.  The fact that the Kawahara traveling
wave ODE is fourth order enables this solution.  It is natural to
conjecture that the TDSW consists of a periodic orbit solution to the
traveling wave ODE that is heteroclinic to an equilibrium,
generalizing homoclinic and heteroclinic solutions studied previously
\cite{haragus_local_2011}.  An open question is the rigorous existence
of a traveling wave solution of the Kawahara equation exhibiting this
structure.  Nonuniformity in the leading edge of the TDSW corresponds
to a forward propagating wavepacket moving with the group velocity for
a distinct wavenumber, approximately that of the periodic orbit. The
uniform wavetrain acts as a channel for the effective dissipation of
energy.  Interestingly, the TDSW only exists for sufficiently large
jumps $\Delta \gtrsim 0.60$, below which waves radiate forward and
backward from the sharp transition region due to the Kawahara solitary
wave band gap.

The generalized Rankine-Hugoniot condition is a kind of nonlinear
resonance condition in the sense that the trailing edge solitary wave
velocity coincides with the adjacent periodic traveling wave velocity.
Such a condition has been assumed previously
\cite{trillo,el2015radiating} but here we show that it is inherent in
the generation of a traveling wave structure within the TDSW.

Although a non-convex dispersion can give rise to TDSWs above
threshold, it is not necessary.  Another model equation, the conduit
equation, also with non-convex dispersion, does not exhibit such
solutions \cite{lowman_dispersive_2013}.  But that model, a
Benjamin-Bona-Mahony type equation \cite{benjamin_model_1972} with
nonlinear dispersion, does display non-classical DSW dynamics at the
DSW harmonic wave edge.  Likely, the principle reason that TDSWs do not
occur is the lack of a linear resonance at the DSW solitary wave edge.

A unique feature of the TDSW is its triple personality.  On the one
hand, it is similar to a dissipative shock wave in that it satisfies a
generalized Rankine-Hugoniot condition and exhibits a steady character
when viewed near the shock front.  On the other hand, the TDSW is
similar to a classical DSW, exhibiting two distinct limits: a small
amplitude, harmonic edge moving with the group velocity and a large
amplitude solitary wave edge moving with the phase velocity.  But the
TDSW is distinct in that the transition from a periodic wave to a
solitary wave occurs almost instantaneously, setting it apart from
DSWs in convex dispersive hydrodynamics and shock waves in dissipative
hydrodynamics.
\section*{Acknowledgments}
The authors are grateful to James Meiss for insightful discussions.

\bibliographystyle{siamplain}
\bibliography{bib}

\end{document}